\documentclass[oneside,10pt]{swissi}

\newcommand{\papertitle}{Multi-Jurisdictional Legal Identity Assurance for Capability Gating}
\newcommand{\papersubtitle}{A Design-Science Proposal for Tiered, Reusable Identity Assurance of Natural, Juridical, and Machine Entities}

\newcommand{\papercitationauthors}{Kurz, W.}

\newcommand{\paperauthors}{%
  Walter Kurz\textsuperscript{1}%
}

\newcommand{\paperaffiliations}{%
  \textsuperscript{1}Swissi Institute for AI,
  \href{mailto:kurz@swissi-ai.institute}{kurz@swissi-ai.institute}%
}

\newcommand{\paperdeclarationaffiliations}{%
  \begin{tabular}[t]{@{}l@{}}
    \textsuperscript{1}Swissi Institute for AI,
    \href{mailto:kurz@swissi-ai.institute}{kurz@swissi-ai.institute}%
  \end{tabular}%
}

\newcommand{\papercorresponding}{%
  Walter Kurz,
  \href{mailto:kurz@swissi-ai.institute}{kurz@swissi-ai.institute},
  \href{https://orcid.org/0009-0006-8045-4775}{ORCID 0009-0006-8045-4775}%
}

\newcommand{\paperfunding}{%
  This research received no external funding.%
}
\newcommand{\paperconflictsofinterest}{%
  The author declares no conflicts of interest.%
}
\newcommand{\paperdataavailability}{%
  The data and code supporting this article are available from the corresponding author on reasonable request.%
}
\newcommand{\paperaitools}{%
  Generative AI tools were used for editorial work only, such as language editing and formatting, in accordance with international scientific standards. The author verified the content and remains fully responsible for the article.%
}
\newcommand{\paperauthorcontributions}{%
  The author is solely responsible for this article.%
}

\newif\ifshowdraftnotes
\showdraftnotesfalse
\newcommand{\gnote}[1]{\ifshowdraftnotes\textcolor{green!50!black}{\sffamily\footnotesize [NOTE: #1]}\fi}
\newif\ifshowroadmap
\showroadmapfalse
\newif\ifshowmodelcard
\showmodelcardfalse

\title{\papertitle \\[0.35em]
\normalsize \papersubtitle}

\author{{\fontsize{8}{10}\selectfont\paperauthors}\\[0.9em]
{\scriptsize\paperaffiliations}}
\date{\vspace{1.5em}}

\begin{document}
\maketitle

\begin{abstract}
Identity assurance is the cost a digital system pays for dishonesty and uncertainty: it exists to make acts attributable when not everyone can be trusted at their word. A common way to pay that cost is flat maximum verification, fixing a single high level of identification and asking each participant to meet it at entry, before any capability is exercised. Paid on everyone, it over-collects, excludes participants who cannot meet a bar they never needed to clear, taxes every interaction with the cost of the rarest high-risk case, and binds the strength of identification to the activity it unlocks. Existing assurance frameworks compound the problem by fixing a small number of per-credential levels inside a single legal space and binding each verification to the institution that performed it, leaving cross-border reuse and the tension between data erasure and evidentiary retention unaddressed. This paper develops, as a design-science proposal, a tiered and reusable model of identity assurance for natural, juridical, and machine entities across jurisdictions. Reading the problem through systems theory, where a system changes only when an entity acts, the model holds the assurance state apart from the capability gate that consumes it, so that identity demand follows the act and the weight of its consequences rather than mere presence: a participant may take part with minimal disclosure, such as a contact handle, and supply more only as an act requires. It comprises a typed entity taxonomy, a two-axis coordinate of disclosed assertion scope and source of information, jurisdiction as a time-indexed attribute of the entity, and reliance recorded as bitemporal, liability-allocated, point-in-time snapshots. Requirements are derived from anti-money-laundering, electronic-identity, and data-protection law, and the proposal is evaluated against flat maximum verification, per-credential level-of-assurance designs, and institutional reusable-KYC reliance.
\end{abstract}


\bigskip

\section{Introduction}

Actors in a system can be dishonest. Many are not, and still act under conditions that only later matter: who acted, when, and under which law. The legal, organisational, and technical systems that govern those acts were not built by any single institution. They grew across time and jurisdiction, and each layer must keep operating while new risks are added to it. Telling honest acts from dishonest ones, and reconstructing the conditions of an act after the fact, is hard and costly.

The system's answer to that difficulty is accountability: it wants to know who did what, when it happened, and under which legal frame the answer must be given. An act can be answered for only if it can be attributed to an actor, and attributed only if the entity behind the act can be identified with enough strength for the consequence at stake. This is why a system identifies at all. Digital identity assurance is the machinery built for that purpose: the procedures by which a relying service can attribute an act to an accountable point, to the strength the act requires, at the moment it requires it.

A common default is flat maximum verification: collecting the strongest identity evidence from every entity at the door, once, before any capability is granted. As a baseline this is coherent and administrable, especially where the capability is high-risk and the law requires customer due diligence.\cite{eu2024amlr,fatf2025recommendations} This paper treats it as the control condition rather than a straw man. As a default, it fails on four counts. It over-collects, gathering more identity data than most interactions require, in tension with data-minimisation and purpose-limitation duties.\cite{eu2016gdpr} It excludes, turning away entities that cannot meet a maximal bar they never needed to clear. It adds friction, pricing every entry at the cost of the rarest high-risk case. It entangles assurance with capability, fixing the strength of identification and the activity it unlocks together rather than matching each to the interaction at hand. What it produces is as flat as what it demands: the outcome is stored as a terminal bit, verified or not, from which nothing about what was checked, by whom, or to what strength can later be queried, replayed, or selectively disclosed. A design that treats every entity as maximal risk at entry exports suspicion from the hardest case to all cases.

A second failure is structural rather than quantitative. Existing assurance frameworks already recognise levels and risk. eIDAS specifies the low, substantial, and high assurance levels for electronic identification schemes, with technical procedures set by implementing regulation, while the European Digital Identity Framework extends that infrastructure around the wallet.\cite{eu2014eidas,eu2015eidaslevels,eu2024eidas2} NIST SP 800-63 separates identity proofing, authentication, and federation, and instructs relying parties to determine whether identity proofing is needed before selecting an assurance level.\cite{nist80063r4,nist80063a4} These frameworks are essential baselines. Their shape is still per-scheme, per-credential, or per-service. The proposed layer targets the architectural object above them: a portable, jurisdiction-aware assurance state that can be relied on across services while preserving the law and policy state that applied at the time of the act.

A third failure is jurisdictional. Electronic communications can form legally relevant acts across borders, and private-international-law instruments determine applicable law act by act in contractual and non-contractual settings.\cite{eu2000ecommerce,uncitral2005electroniccommunications,eu2008romei,eu2007romeii} What fails is the architecture around the anchor. The framework is fixed at enrolment, with no change event and no timeline, so an account stays hard-bound to the jurisdiction in which it was created while the life behind it moves. Identity-assurance frameworks harmonise schemes and credentials, but they do not by themselves supply a portable assurance state whose jurisdictional anchor changes over time above those schemes.\cite{eu2014eidas,eu2015eidaslevels,eu2024eidas2,nist80063r4} The person who moves is left with a dichotomy the frameworks do not register: abandon the account and re-enrol as a stranger, losing the record, or keep using it and silently fall out of compliance with the regime that now actually governs. Treating jurisdiction as a time-indexed attribute of the entity rather than as a property of the account is what this dichotomy demands, and it is the part of the proposal the bitemporal record carries.

This paper reads these failures through systems theory, a deliberate analytical commitment rather than a borrowed vocabulary: they are clearer as system properties than as a list of implementation defects. Read through Niklas Luhmann's account of social systems as communication, the system does not know a person; it registers acts and conditions what may follow from them.\cite{luhmann1995social} Rule accretion can then be read as autopoietic reproduction, the concept Maturana and Varela introduced for living systems and Luhmann carried into social theory, rather than administrative carelessness.\cite{maturana1980autopoiesis,luhmann1995social} The system reproduces itself from its own distinctions, and answering each new harm with a new prohibition is one way it grows, which is why more rules do not converge on less complexity. The lens reframes the design question: not how much identity to collect, but why a system must identify at all, and how much identification a given act actually warrants.

An enrolled entity is not yet an actor in the sense used here.\cite{luhmann1995social} It becomes an actor only when it acts: when a natural person acting through a machine concludes a contract with a juridical entity, when a representative binds an organisation, or when a service grants a capability in reliance on a presented state. A registry of five million dormant entities raises no legally relevant event. The system changes with the act. Identity demand attaches to the act and to the weight of its consequences, not to mere enrolment. Flat maximum verification inverts this, pricing identity at enrolment rather than at the act, which is the systemic origin of its over-collection.

The same lens explains why the problem becomes acute in mediated communication. In face-to-face interaction the source of a communication and the body that carries it arrive together, so embodied co-presence supplies identification as a side effect.\cite{goffman1967interaction} Each dissemination medium loosens that bond, and digital communication can sever it: the sender's identifier may be absent, forged, delegated, automated, or merely asserted, and the context that once travelled with the speaker is reconstructed only if the system requires it. Identity assurance is the explicit reconstruction of what co-presence once performed without a separate procedure. And what co-presence once supplied for free, the model developed below prices as a distinction of its own: the same fact confirmed over a remote channel and confirmed in physical presence are different classes of evidence.

A new kind of participant now enters this mediated communication. A machine trained on the recorded products of human communication, the objective contents Popper called World 3, can interact with them rather than merely store them, producing system-relevant consequences without being a source of intent, close to what Elena Esposito calls artificial communication.\cite{popper1972objective,esposito2022artificial} An identity model for digital communication must include machine entities while anchoring accountability for their acts in a natural or juridical entity capable of bearing consequence. The entity taxonomy below, and paper 17, set out how that borrowed will is attached.

The flat default and the per-credential frameworks fail against a single shared measure, which the model developed below derives from its limiting cases. Treat complete honesty and complete deception as idealisations, the way the exact sciences reason from a frictionless plane: at the deception-free pole the identity an act demands falls to a bare identifier, and at the all-deception pole it rises to flat maximum verification. Real interactions lie between, and where a given one falls depends on the act, on its exposure to deception and the weight of its consequences. Flat maximum verification is the design for one pole only; the design developed here follows the relationship instead of clamping it, beginning at the identifier floor and rising only as far as each act requires.

The same limit shows why assurance and capability belong on separate axes. At the border of a deception-free society a traveller is believed on her word, yet she may still be refused entry: verification has gone to zero while the decision to admit or refuse remains. Whether an act is permitted is independent of how strongly its actor is identified, and the model keeps the two apart throughout.

The gap is architectural. Existing assurance frameworks define important assurance categories, but they stop short of a progressive, proportionate, jurisdiction-aware assurance layer that is separate from the capability it gates. The missing object is reusable across relying parties and borders, discloses by named part rather than as a packed record, allocates liability for reliance, preserves the legal and policy frame in force at the time of each act, and remains defined across natural, juridical, and machine entities. This is the part that a wallet, a credential scheme, or a local onboarding policy does not by itself provide.

This paper develops that missing layer as a design-science proposal, treating the assurance model as an information-systems artifact to be specified and evaluated against alternatives.\cite{hevner2004designscience} It models identity assurance for natural, juridical, and machine entities across jurisdictions, grades the assurance state on two axes, what an entity has disclosed and who stands behind each disclosed item, holds that state apart from the capability gate that consumes it, and records each act of reliance as a point-in-time, liability-allocated snapshot that a later audit can replay under the law and policy version in force at the time. The proposal is a reusable assurance layer whose state can be carried between relying parties rather than re-collected. It is evaluated against flat maximum verification, per-credential level-of-assurance designs, and institutional reusable-KYC reliance under requirements derived from anti-money-laundering, electronic-identity, and data-protection law.

\section{Related Work}

Formal assurance frameworks define the credential-level baselines. eIDAS requires notified electronic-identification schemes to specify assurance levels low, substantial, and high, with minimum technical specifications set by Commission Implementing Regulation 2015/1502, and the European Digital Identity Framework extends that infrastructure around the European Digital Identity Wallet, specified by its architecture and reference framework and built as an open reference implementation.\cite{eu2014eidas,eu2015eidaslevels,eu2024eidas2,eudiarf2026,eudiwallet2026ios} These levels grade an identification means inside one scheme and one legal space, and do not hold a reusable assurance state apart from the capability that consumes it. The wallet is the strongest single corroborator an assurance state can absorb, yet in the design proposed here a sovereign scheme accelerates assurance without ever being a precondition for it: a rule developed below keeps assurance reachable through non-sovereign confirmation where a jurisdiction offers notarial, public-officer, qualified-trust-service, or functionally equivalent confirmation channels.\cite{hcch1961apostille,uncitral2022mlit} Sovereign schemes are advancing on these terms, from the European wallet to Switzerland's state-issued e-ID, whose narrow approval after an earlier rejection registers the public concern over exclusion and over-identification that the equity rule is meant to contain.\cite{ch2024eid,eidas2concerns2023} NIST SP~800-63 comes closest to separating the dimensions: it decomposes assurance into identity proofing, authentication, and federation, and directs the relying party to determine whether proofing is needed before selecting a level.\cite{nist80063r4,nist80063a4} Federation is the part that most resembles reuse, since it lets a relying party consume an identity assertion it did not itself proof; it passes that assertion at authentication time inside one pre-agreed trust framework, rather than holding a portable, holder-presented assurance state apart from the gate, carrying it across legal contexts, or recording reliance as an allocated, replayable point-in-time event. This paper targets that object: a reusable, jurisdiction-aware assurance state that later capability gates consume across legal contexts.

Organisational identity has its own baselines, and they stop short of the same object. The Legal Entity Identifier, standardised as ISO~17442 and administered by GLEIF, gives every legal person a register-anchored global identifier with verified reference data\cite{gleif2025lei}; it identifies the entity but grades no assurance and gates no delegation. The verifiable LEI comes closest: it issues the identifier as a holder-presented credential and binds officer roles to the entity through role credentials\cite{gleif2024vlei}, yet it asserts a role rather than bounding the entity's assurance by the assurance of the natural persons accountable for it. The legal analogue of that bound appears in the beneficial-ownership regime, which traces legal persons and legal arrangements to natural persons who ultimately own or control them.\cite{fatf2023beneficialownershiplegalpersons,fatf2024beneficialownershiplegalarrangements,eu2024amlr} The artifact developed here turns that pressure into a composite gate in which the entity's assurance is capped by its representatives' assurance and the validity of their delegation. That bound, developed below as the inheritance bound, is the organisational part of the gap.

Institutional reusable KYC is the reuse comparator in this paper: an arrangement in which participating institutions exchange due-diligence data, documents, completed checks, or standardized profiles under retained relying-party responsibility. FATF Recommendation~17 lets an obliged entity rely on a third party for specified customer-due-diligence measures while ultimate responsibility for those measures stays with the relying institution.\cite{fatf2025recommendations} The current EU AML directive permits retained-responsibility third-party reliance, while the 2024 AML package recasts AML supervision and due-diligence architecture for the incoming single-rulebook framework.\cite[arts.~25--27]{eu2015amld4}\cite{eu2024amld6,eu2024amlr} SWIFT's KYC Registry standardises the exchange of due-diligence data and documents among participating financial institutions.\cite{swift2026kycregistry} The Nordic banks' consortium utility, Invidem, pooled corporate KYC into a shared clearing house on the same pattern.\cite{biia2020invidem} Reuse is already valuable but remains pinned to institutions and to retained-responsibility reliance; this paper moves the object to a holder-carried assurance history whose every reliance is an allocated, point-in-time record.

Holder-controlled identity standards supply the adjacent credential substrate. The W3C Verifiable Credentials data model defines an issuer-holder-verifier pattern in which holders present verifiable credentials to verifiers.\cite{w3c2025vcdm} Self-sovereign-identity literature links that pattern to selective disclosure, data minimisation, and user control over disclosure granularity.\cite{schardong2022ssi} Those works address credential representation and presentation. The contribution here is the legal-assurance layer that specifies entity type, jurisdictional anchor, source of information, gate predicate, and reliance state around the credential.

The resulting gap is architectural. Prior work supplies assurance levels, wallet infrastructure, organisational identifiers, reusable institutional KYC, and holder-presented credentials, each strong in its own dimension. The missing object joins them: a progressive, jurisdiction-aware assurance state separated from the capability it gates, defined across natural, juridical, and machine entities, bounding a legal person's assurance by the assurance of the accountable persons behind it, and replayable as a liability-allocated point-in-time reliance event.

\section{Contribution}
This paper proposes a design-science contribution of two kinds, a construct and an artifact.\cite{hevner2004designscience} The construct is a typed taxonomy of the assurance subject: the subject is an entity rather than a customer, credential holder, account, or relying-party user, and an entity becomes an actor only in the act a gate evaluates. The taxonomy separates natural, juridical, and machine entities, with legal arrangements as a personality-less sibling, because each type carries a different imputation regime and a different evidence problem. Within the juridical type it proposes an inheritance bound: a legal person's assurance is capped by the assurance of the accountable natural persons behind it and the validity of their delegation, a composite that a register-anchored identifier or role credential such as the LEI or vLEI identifies but does not bound.\cite{gleif2025lei,gleif2024vlei}

The artifact is the ontology package that operationalises the construct, and its organising property is the separation of assurance from capability. The package holds a reusable assurance state, the assertions an entity has accumulated and the sources that confirm them, apart from the capability gate that consumes it, so that the same state can pass a low-disclosure gate, fail a higher one, and later be replayed for audit. Existing frameworks fix the strength of identification and the activity it unlocks together; the artifact proposed here keeps the two on separate axes, which is what lets a single assurance state be reused across services and gates rather than re-collected at each.

Within the artifact, an assurance coordinate composes two axes. The assertion axis records what identity-relevant claim is disclosed into a capability context; the source-of-information axis records who or what stands behind that claim. This separates disclosure scope from evidentiary strength, so that a low-disclosure state supports low-consequence participation while a higher-consequence gate requires stronger assertions, stronger sources, fresher evidence, or a statutory mapping such as customer due diligence.\cite{eu2024amlr} Where per-credential level-of-assurance designs grade an identification means as a single rung, the coordinate grades disclosure and confirmation independently. The number of levels on either axis is a design parameter rather than a claim: the contribution is the two-axis structure, not a particular rung count.

The fourth contribution is reusable, jurisdiction-aware reliance. Jurisdiction is modelled not as a fixed property of an account but as a time-indexed attribute of the entity: parallel time-tracked memberships with one active anchor, where facts are stored once and jurisdiction-relative predicates are evaluated under the governing law at check time, and where a sovereign scheme accelerates the climb without ever being a precondition for it. Existing level-of-assurance frameworks define scheme or proofing levels\cite{eu2014eidas,nist80063a4}, KYC utilities reuse documentation inside institutional arrangements\cite{swift2026kycregistry,biia2020invidem}, and credential standards let holders present verifiable claims\cite{w3c2025vcdm}; the artifact combines these separable functions into a bitemporal assurance state in which each reliance event records the claim state, valid time, record time, jurisdictional anchor, governing law, requirement version, parties, and liability boundary, so that a past act can be replayed under the law and policy in force when it occurred.

Each of these is offered as a design proposed for evaluation against alternatives, not as the only admissible one: the claim is the architecture and its separations, not any particular setting of its levels or tiers.

\section{Research}

\subsection{Identification thresholds for capability-bearing interactions}


Identity assurance is modelled here as a socio-technical attribution system: an ordered environment in which acting units, evidence sources, relying services, regulators, and jurisdictional rules interact. The system boundary is drawn around the identity-relevant interaction, rather than around a database, wallet, or single provider. Within that boundary, the design problem is to produce and reuse an assurance state that can be evaluated when a capability is requested.

The system sits between actors and capabilities. It observes an actor through presented evidence, records the source and time structure of that evidence, attributes the actor to a jurisdictional frame, and lets a declared gate consume the resulting state. The core question is who or what may do which act, under which law, on which evidence, at which time, and with which relying party accepting which residual responsibility.

The design premise is capability-proportional participation. Within the applicable legal frame, a low-assurance state is adequate for low-consequence interaction, while higher-consequence acts require stronger evidence, fresher confirmation, or an explicit regulatory mapping. This is the paper's non-KYC lower floor: legitimate participation with minimal identity payload, constrained by capability rather than treated as failed verification. A request for additional information is justified by a gate predicate, a legal basis, or a relying-party risk decision, and the consequence of an unmet predicate is scoped to the requested capability. Reusable assurance turns verification into carried evidence: a verified claim can be presented again with consent, source, time, and scope, while relying parties apply their own gates and retention duties. The design shifts repeated identity collection toward auditable reliance on a bitemporal assurance state.


The limiting case makes the burden visible. Call a society deception-free if no participant lies, steals, or misuses what it learns. The case is not omniscient: participants can still be wrong, forget, collide, and die. In such a society the identity tax falls almost to nothing. A traveller names herself at a border and is assessed on that claim, with no passport, password, or proof, because the claim and its truth do not diverge.

The instructive part is what remains. Identity does not fall to zero; it falls to the bare identifier. A name, a date and place of birth, or a stable reference such as a national number is still needed to distinguish one person from another, address the right party, and keep a record straight when honest participants misremember. Documentation survives for the same reason: it carries information from one party to the next and from one time to another. The deception-free limit separates two things ordinary practice fuses: the identifier, which coordinates, and the confirmation, which guards against deception. The first survives perfect honesty; the second is the tax.

Even perfect honesty leaves a residue an actor cannot supply about itself. Whether it still exists, whether it has legal capacity, and whether another party has granted it authority are facts that no honest declaration can establish, because the actor cannot be their source. A deception-free society still needs external confirmation for such facts, which is why the self is the floor of the confirmation scale and never its guarantee.

We treat the two pure cases, complete honesty and complete deception, as limiting idealisations, the way the exact sciences reason from a frictionless plane or an ideal body: neither extreme occurs in pure form, but the two expose the law that governs everything between them. At the deception-free pole, demanded assurance falls to the identifier floor; at the opposite pole, where any party may be lying, it rises to flat maximum verification. Real interactions lie between, and where a given one falls depends on the act: on how much deception it is exposed to and on the weight of its consequences. Demanded assurance can then be written as a function of those two quantities,
\begin{equation}\label{eq:assurance}
  A \;\approx\; f(\delta \cdot \kappa),
\end{equation}
where $A$ is the assurance an act demands, $\delta$ its exposure to deception, and $\kappa$ the weight of its consequences. Flat maximum verification is the right design at one pole only: it fixes $A$ at its maximum and charges the all-deception price on every act, including those whose deception exposure or consequence is small. The design developed here follows the function instead of clamping it, beginning at the identifier floor and climbing only as far as each act's exposure and consequence require.

The label \enquote{know your customer} names the apparatus after a relationship (the customer) and an epistemic state (knowing), and neither carries the legal weight. The weight sits elsewhere: an act occurs, it has consequences, and the legal system must impute that act to a point it can hold responsible. This paper reads Kelsen and Luhmann as supporting that imputation view: legal personhood functions as a \emph{Zurechnungspunkt}, a point of imputation, the place at which rights, duties, and the consequences of acts are made to land.\cite{kelsen1967pure,luhmann1995social,luhmann2004law} Natural and juridical persons are two kinds of such point. The reframing restates the elementary fact that an actor acts and the act has consequences in the terms the legal system already uses for itself.

\enquote{Actor} is on this reading not a softer word for \enquote{user} but the name for that to which a consequential act is imputed. The identity apparatus, customer due diligence included, exists to make imputation defensible: to ensure that when an act with legal effect occurs there is an accountable point behind it and the evidence to bind it there. Identity assurance is, in this sense, infrastructure for imputation, and the \enquote{know your customer} construction names the byproduct (knowing a counterparty) rather than the function (imputing an act).

This recovers the systems-theoretic reading. Social systems, in Luhmann's account, are composed of communications and actions rather than of persons, and the human being belongs to the environment of the system. The system does not \enquote{know the customer}; it registers acts and conditions them. \enquote{Know your customer} is then a category error twice over: it places a person where the system has only actions, and a relationship where the system has only imputation. An actor-centred framing restores both, so that the unit is the act, the actor is its point of imputation, and assurance is the strength with which the system can stand behind directing a consequence to that point. The principle on which the question of why a system identifies at all then turns is the following.

\begin{quote}
A system demands identity only to the degree that an action's consequences must be imputable to an accountable point, and only in proportion to the weight of those consequences.
\end{quote}

The principle has a converse that completes it. At the border of the deception-free society the traveller is believed on her word, yet she may still be refused entry: demanded assurance has fallen to the identifier floor while the decision to admit or refuse stands untouched. How strongly an actor is identified and whether its act is permitted are independent quantities. A design that derives one by setting the other has fused two different questions. This is the second design commitment: the assurance state is held apart from the capability gate that consumes it, so that the same state can pass one gate, fail another, and later be replayed for an audit, without being re-collected for any of them.

The deception-free limit yields one further structural result, and it is the one the rest of this paper builds on. The limit split identity into two components that ordinary practice fuses: the identifier, which coordinates, and the confirmation, which guards. The two answer different questions and vary independently. Which facts about its actor an act must be able to read is fixed by the act's function and its legal mapping: an age-gated purchase needs a date of birth and nothing else, a payout needs a payment route, a conveyance needs the civil name on the deed. How strongly each fact must be guarded is fixed by the act's exposure and weight, the two arguments of \cref{eq:assurance}. A model that grades identity on a single scale has no place to keep this difference: it must answer with one number where every act asks two questions, which facts, and how strongly each. The assurance state needs two orthogonal axes, an \emph{assertion} axis recording what an entity has disclosed, and a \emph{source-of-information} axis recording who stands behind each disclosed item. What an act demands is a set of paired requirements,
\begin{equation}\label{eq:demand}
  D(a) \;=\; \bigl\{\, \bigl(r,\; s_{\min}(a,r)\bigr) \;:\; r \in R(a) \,\bigr\},
\end{equation}
where $R(a)$ is the set of assertion classes the act's function and legal mapping require, and $s_{\min}(a,r)$ is the minimal confirmation strength demanded for each, rising with the act's deception exposure $\delta$ and consequence weight $\kappa$ from \cref{eq:assurance}. Flat maximum verification is the degenerate case in which $R(a)$ is everything and $s_{\min}$ maximal for every act; the per-credential level designs are the case in which $D(a)$ must be squeezed through one shared index. The subsections that follow construct what this demand set presupposes: a taxonomy fixing which catalogue of assertion classes an entity carries, a jurisdictional anchor fixing under which law a demand is evaluated, the two axes in full, the grid they span with the coordinate that names its cells, and the gates that consume them.

Everything downstream, the entity taxonomy, the assertion and source axes, the grid and its coordinate, the capability gates, and reusable reliance, is machinery for producing imputation at the right strength, at the right time, and for the right actor.

\subsection{Entity taxonomy for assurance evaluation}
The assurance subject is typed before it is graded. The section uses two terms strictly. An \emph{entity} is the static bearer of an assurance state, present because it can be held or because it can actuate. An \emph{actor} is that entity in the act to which a consequence is imputed. The distinction matters because technical action and legal identity need not coincide. A machine may change the state of a system and must be modelled as an entity for that reason; present law still channels obligations and liability through natural or juridical persons around the machine, such as providers, deployers, manufacturers, and other economic operators.\cite[arts.~3, 16, 25, and 26]{eu2024aiact}\cite[art.~4]{eu2024pld} The taxonomy is fixed by that purpose: a system identifies an entity so that, should harm follow, it can direct consequence to an actor's accountable point, while disclosing nothing until an act requires it. To answer for an act has an operational meaning. An act is answered for when its consequence can be addressed to a point the legal order can hold, that is, sue. The taxonomy turns on one question asked of every participant, and a second asked only of those that pass the first.

The first question is whether the participant is \emph{suable}: whether the law will let an act or an asset be addressed to it in its own name. The second, asked of the suable, is the \emph{source of behaviour}: whether it can originate an act or is acted through. The two questions partition the space. A suable participant that originates behaviour is a \emph{natural} entity (N), a natural will that both acts and can be held. A suable participant that originates no behaviour of its own, but is acted through, is a \emph{juridical} entity (J), a construct the law holds although action reaches it through organs, representatives, trustees, or mandated machines. A participant that originates behaviour while lacking suability is a \emph{machine} entity (M), an engineered actuator that changes the system but cannot stand as the legal endpoint of responsibility. In legally relevant operation it must attach to a suable N or J, or to a permitted constellation that supplies that endpoint. The remaining logical combination, a participant that neither acts nor can be held, is an inert object outside the assurance model. The accountability axes yield exactly three entity types, differing in kind rather than degree.

Two qualifications keep the partition exact. First, the machine type covers engineered or programmed actuators acting under a mandate the system can read. Mere causal forces are treated as instruments or objects in an accountability path, according to the legal relation that brings them into the act. Second, suability has two grades, and only the formal grade types the entity. A participant is \emph{formally} suable when it can be named as defendant, and \emph{substantively} suable when there is in fact a will, estate, or legally addressable accountability surface to hold. The type cut uses formal suability; substantive suability governs the artifact's admission state. A juridical shell with no traceable control point, and a machine with no accountable attachment, may be formally addressable in some settings yet substantively empty for the assurance system; the model holds them dormant within their type. Dormancy is an admission state, not a fourth type.

\begin{figure}[!htbp]
\centering
\begin{tikzpicture}[
  rootbox/.style={draw, rounded corners=3pt, minimum height=0.85cm, minimum width=4.4cm,
                  align=center, font=\scriptsize},
  headbox/.style={draw, rounded corners=3pt, fill=black!4, align=center,
                  font=\scriptsize, text width=3.4cm, minimum height=1.05cm, inner sep=2.5pt},
  notebox/.style={draw, dashed, rounded corners=2pt, align=left,
                  font=\fontsize{6}{7}\selectfont, text width=3.4cm, minimum height=1.75cm, inner sep=3pt},
  conn/.style={semithick},
]
\node[rootbox] (root) at (0,0) {Entity\\[-0.25em]{\fontsize{6}{7}\selectfont addressable participant in an accountability path}};
\node[headbox] (N) at (-5.2,-1.7) {Natural\\N};
\node[headbox] (J) at (0,-1.7)    {Juridical\\J};
\node[headbox] (M) at (5.2,-1.7)  {Machine\\M};
\node[notebox] (cN) at (-5.2,-3.5) {suable; originates behaviour. Assertion catalogue N0--N10 $\times$ confirmation S1--S7; a tier is derived from the filled grid, never a row of it.};
\node[notebox] (cJ) at (0,-3.5) {suable; behaved-through. One catalogue, register-parameterised: company, association, foundation, public body, partnership; a legal arrangement (trust) is the flat trustee-imputed corner.};
\node[notebox] (cM) at (5.2,-3.5) {lacks suability; actuates; must attach. Descriptors: capability (M1/M2/M3), trust (vector), control. Headline kinds: bot, smart contract, ai, true~ai.};
\coordinate (br) at (0,-0.8);
\draw[conn] (root.south) -- (br);
\draw[conn] (br -| N.north) -- (br -| M.north);
\foreach \t in {N,J,M} {
  \draw[conn] (br -| \t.north) -- (\t.north);
  \draw[conn] (\t.south) -- (c\t.north);
}
\end{tikzpicture}
\caption{Entity taxonomy for assurance evaluation. Two accountability cuts, suability then source of behaviour, yield three entity types, natural (N), juridical (J), and machine (M); the remaining logical branch lacks both suability and behaviour and is an inert object outside the model. The text develops the type definitions, the juridical catalogue and its trustee-imputed arrangement corner, the machine descriptors, and the grid from which a tier is derived.}
\label{fig:entity-taxonomy}
\end{figure}
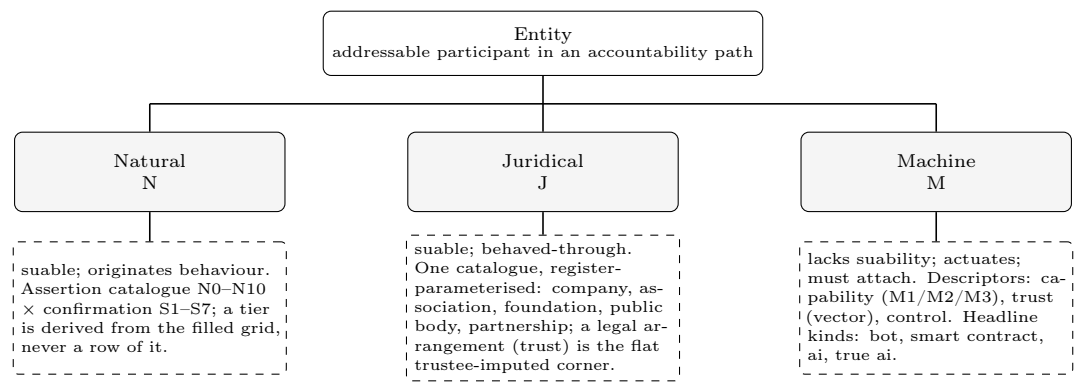

The resulting taxonomy, drawn in \cref{fig:entity-taxonomy}, renders this two-cut derivation together with the three decisions that most distinguish the model from a conventional identity ontology. It presents the three types as the outcome of the suability and behaviour-source cuts rather than as a borrowed list of legal categories; it draws the juridical type as a single catalogue whose company, association, foundation, public-body and partnership forms are values of one register parameter rather than separate branches, with the legal arrangement as its flat, trustee-imputed corner; and it replaces a machine subtree with three descriptor axes, capability, trust, and control, under which the familiar bot, smart-contract and artificial-intelligence labels are regions rather than kinds. Read downward, each type resolves into a grid of disclosed assertion scope against source of information, from which a tier is derived. The figure is in this sense a map of where consequence can land and how strongly each landing point is evidenced, not a ladder of ranks.

With this vocabulary fixed, the figure should be read as an imputation map rather than a population list. Natural persons, juridical entities including the legal arrangements discussed below, and machines acting under mandate can all carry assurance state. They become actors only in an act. The entity is what can be classified, evidenced, gated, and later audited; the actor is the entity as the system directs consequence to it.

Two consequences follow. First, entity type is an imputation regime rather than a classificatory bucket: what a natural person can be held to, what counts as its identity, and what evidence binds it differ from a juridical entity, and differ again for a legal arrangement that sits within the juridical type as a distinct corner, while none of it applies to a machine, which can actuate but cannot be the legal root point of imputation. The entity taxonomy is the structure of imputation itself, which is why it is load-bearing and why it precedes the assurance model: it is the map of where consequence can land.

Entity types differ not only in how they are identified but in whether they can originate action at all, and this asymmetry, the second cut, shapes both the threat model and the structure of accountability. Natural and machine entities are sources of behaviour: a natural person acts through a body and a will, and a machine executes actions and may do so adversarially, whether through autonomy, misalignment, or capture. A juridical entity, the legal arrangements folded into it included, has no behaviour of its own; it acts only through the natural persons that serve as its organs or trustees, or through machines acting under its mandate. The distinction is categorical: a natural or machine entity behaves, whereas a juridical entity is behaved-through.

We propose, on this basis, that malice, understood as the conjunction of behaviour and intent, originates in a behaviour-bearing entity. A juridical entity is the legal locus to which another behaviour source is imputed: a natural person's act, or a mandated machine's act together with the natural or juridical point that controls it. What is colloquially termed a malicious entity is, structurally, the malice of natural persons acting through the entity and imputed to it, or the legally allocated consequence of a machine act under mandate. Common-law corporate attribution doctrine expresses the same asymmetry in one doctrinal setting when it constructs culpable corporate states by attributing a natural person's intent to the entity through the directing-mind, or identification, doctrine.\cite{tesco1971nattrass} In this model, juridical fault is an imputed state rather than an original behaviour source.

This asymmetry determines what each entity type must be verified against. A natural entity is verified against impersonation of a living, willed body, since the threat is a false will presented as genuine. A machine entity is verified against action outside its mandate and against capture, since the threat is behaviour that exceeds or betrays its authority. A juridical entity is verified against impersonation of the registered entity and against the integrity of the natural persons accountable for it, since the entity itself can neither will nor offend; the threat is that an impostor presents as the entity, or that the natural persons behind it use it as a vehicle.

A juridical entity with no accountable behaviour-source behind it is an origin-less construct for this artifact: a point of imputation with nothing to impute from. The system holds such an entity dormant. The attachment requirement is a design rule derived from imputation and beneficial-ownership pressure: an active juridical entity must present a traceable ownership, control, organ, trustee, or representative path that reaches at least one identified natural person or a legally recognised officeholder acting in fiduciary capacity. Intermediate links in the chain may be other juridical entities, as in ordinary holding structures, but the transitive closure of the submitted assurance path must contain a natural will before the entity can receive an active assurance state. A chain of constructs that terminates in no accountable natural or fiduciary point is the structure the rule excludes, and the structure that opaque ownership exploits.

This requirement is narrower than, and informed by, beneficial-ownership law. Anti-money-laundering sources require transparency of legal persons and legal arrangements so that authorities and obliged entities can identify natural persons who ultimately own or control them.\cite[arts.~51--63]{eu2024amlr}\cite[recs.~24--25]{fatf2025recommendations}\cite{fatf2023beneficialownershiplegalpersons,fatf2024beneficialownershiplegalarrangements} That regime is ordinarily presented as a compliance obligation. The imputation reading supplies the artifact rule: because natural persons originate action, accountability for a juridical entity's assurance state should ground out in identified natural persons or fiduciary officeholders whose authority can be evidenced. The contribution here is to translate that legal pressure into an admission and gate rule, not to claim that present law already states the rule in this exact form.

The same principle, stated as a quantity rather than a condition, yields the inheritance bound developed later: a juridical entity's assurance cannot exceed the assurance of the natural persons accountable for it, and a composite gate over a juridical entity reads the entity's state, the attached natural persons' states, and the validity of the delegation between them together. The attachment rule establishes that the accountable natural persons exist and are identified; the inheritance bound establishes how their assurance propagates to the entity. Because more than one natural person is ordinarily accountable, a board, a set of partners, a body of controllers, the attachment is to a set rather than to a single anchor, and the bound is evaluated per act over the subset the act's delegation actually traverses: the assurance of a jointly required signing is that of its weakest required signer, while where any one of several may act the entity may put forward its strongest. The set carries two readings the gate keeps apart, the decision rule that selects the authorising subset and so the bound, and the liability mode that fixes the recovery surface.

Stated formally, with $\tau(\cdot)$ the tier a gate derives from an entity's filled grid (\cref{sec:coordinate}) and $\mathcal{Q}(a)$ the family of subsets of accountable natural persons whose delegation validly authorises act $a$, the inheritance bound reads
\begin{equation}\label{eq:bound}
  \tau(J \mid a) \;\le\; \max_{Q \in \mathcal{Q}(a)} \; \min_{N \in Q} \; \tau(N),
\end{equation}
the inner minimum expressing that a jointly required signing is as strong as its weakest required signer, the outer maximum that an entity may put forward its strongest admissible constellation where any one of several may act. The bound is evaluated per act, over the delegation the act actually traverses, and it presupposes the attachment rule: $\mathcal{Q}(a)$ must be non-empty for the entity to act at all.

Within a type, an entity is graded on two axes the type selects but does not change: the disclosed \emph{assertion scope}, the catalogue of identity items it can reveal, and the \emph{source of information}, the accountable weight behind each disclosed item. The catalogue differs in length by type, rich for a natural entity, register-bound for a juridical one, a short descriptor set for a machine, while the source axis is shared. Both axes, the grid they span, and the coordinate that names a cell are developed in the assertion-granularity, source-of-information, and coordinate subsections below; here it is enough that the type fixes which catalogue an entity carries and the act fixes which of its cells must be read. What is loosely called a tier is a value derived from the filled grid against a threshold, never a row of it, which is why \cref{fig:entity-taxonomy} shows a grid rather than a ladder.

The juridical type looks, in the figure, almost as flat as the natural one, and a reader will reasonably object that the law plainly knows many kinds of legal person, a company, an association, a foundation, a public body, a partnership, a trust, each of which appears to deserve its own treatment. They do differ, but not where the model types. A distinction earns a separate type only when it changes the assertion catalogue or the behaviour of the confirmation axis, and the registered juridical forms change neither. They share one catalogue shape, legal name, founding instrument, register entry, organs or representatives, purpose, status, and the accountable natural persons behind them; they share the one source scale; and they share a single imputation regime, a constructed point the law holds and that grounds out, through attachment, in natural persons. What varies among them is which authoritative register answers the source axis, a commercial register for a company, a supervisory authority for a foundation, a public-law instrument for a public body, none at all for an unregistered association, together with a small set of predicates a gate reads directly: whether the form carries separate personality, whether beneficial-ownership tracing applies, and where imputation lands.\cite{eu2017companylaw,fatf2023beneficialownershiplegalpersons,fatf2024beneficialownershiplegalarrangements} Drawing each form as its own subtype would multiply catalogues identical in shape and preserve differences that no gate reads from the type. The model keeps them as juridical-form predicates for the same reason it declines to over-collect: its machinery should track only the distinctions an act actually consumes.

The differences move from the type tree, where they would harden into separate machinery, to a predicate set a gate can interrogate one at a time. The two genuine variations of the regime show how. A partnership is registered and graded like a company where the governing law supplies a register, yet the model can route personal liability to partners through \texttt{separate-personality=false} where that is the legal effect. A legal arrangement such as a trust is treated through the fiduciary or trustee to whom imputation falls, with beneficial-ownership and transparency rules handled as legal-arrangement predicates rather than as a new entity type.\cite{fatf2024beneficialownershiplegalarrangements} The outer boundary of the type follows the same functional test: a formation is juridical when the law makes it suable in its own name, or through a required officeholder or fiduciary capacity, with an estate or patrimony distinct from its participants. Where the law instead reaches through to participants, as with an informal group, the assurance subject is a constellation of the natural persons themselves. A register is strong evidence that this line has been crossed, never the definition of it.

A machine entity carries no catalogue of life-facts; it is described instead by three orthogonal axes a gate reads together. Its \emph{capability} is the scope of action its construction permits, ordered as a deterministic actuator whose decision path contains no model inference (M1), a model-driven actuator acting inside a human-authored mandate (M2), and a reserved class for a future actuator that sets goals beyond a human-authored mandate (M3). Its \emph{trust} is a fixed, versioned vector of confirmation dimensions, provenance, integrity of the running instance against its attested artifact, verification and audit status, explainability, and data sovereignty, each read by the gate that depends on it rather than averaged into one figure. Its \emph{control} records how the actuator can be reined in: immutable, upgradeable, or governed, the custody of any administrative key, and the recovery surface available when it misfires. The familiar labels, a bot, a smart contract, an artificial-intelligence agent, are regions of this descriptor space rather than types: a bot and a smart contract are both deterministic (M1), separating only on integrity and control, an on-ledger contract verifiable in its bytecode and executing without an administrative custodian where a bot is operator-attested and operator-held. Automated contracting instruments confirm that automated systems may participate in legally relevant contracting processes, while accountability is still resolved through legal actors around the system.\cite{uncitral2024automatedcontracting} The labels form a machine headline, the analogue of the natural and juridical classes, and like them are never a gate input. Where law supplies an authoritative classification it maps onto these axes: the European Union's artificial-intelligence risk class is a ceiling on permitted capability, while its conformity assessment and notified-body attestation are a source of confirmation for the trust vector.\cite[arts.~6, 43]{eu2024aiact} The legal endpoint still remains outside M. Contemporary EU instruments regulate the system by allocating obligations and liability to legal actors around it, and the European Parliament has expressly treated civil liability for AI systems as requiring accountable persons rather than legal personality for the system itself.\cite[para.~7]{ep2020ailiability} The model follows the same split: M may act technically, but a legally relevant M-operation is admitted through N+M, J+N+M, or a restricted J+M path only where the governing legal framework supplies direct juridical responsibility, recovery, and compensation.

A smart contract is the canonical machine entity: deterministic on-ledger code addressed by its own address, holding assets and producing effects with no will of its own. A decentralised autonomous organisation is one application of such a contract, and it shows the type cut at work. A registered DAO, in a jurisdiction that gives such an organisation a statutory wrapper, supplies a suable construct with its own estate and is a juridical entity acting through a machine, with J+M completed by an accountable natural anchor as J+N+M for consequential operation.\cite{wyoming2021daosupplement,wyoming2024duna} A wrapperless DAO raises different questions; enforcement and litigation practice can reach through to the participants or treat the association through the available procedural category.\cite{cftc2023ookidao} The model records that edge case as a machine attached to the set of natural persons who control it, $\mathrm{M}+\{\mathrm{N}\dots\}$, served through the contract address and answerable through those controllers where the governing law permits that route. The example is a design test for the taxonomy, not a claim that all jurisdictions resolve DAOs alike.

The attachment requirement is indifferent to substrate as an artifact rule. A chain of control terminating in no natural person or fiduciary accountability point is origin-less for assurance purposes whether it is built of juridical shells or autonomous machines, and the model holds it dormant in either case. Beneficial-ownership law supplies the legal pressure for juridical chains; AI and automated-contracting sources supply the legal pressure for machine-mediated acts.\cite{fatf2023beneficialownershiplegalpersons,fatf2024beneficialownershiplegalarrangements,eu2024aiact,uncitral2024automatedcontracting} Where a machine can act beyond a scope fixed in advance, a model-driven or goal-setting machine (M2 or M3), live control or live recovery is required before it is granted active capability, because its behaviour is bounded through oversight rather than through fixed code. A deterministic and immutable machine (M1) is the single case that may act with no present means of halting it, since its whole behaviour is fixed at admission and a recovery or liability surface can be sized in advance to the harm that bounded behaviour can cause. Live control supplies what immutability already provides, a behaviour boundary before the act, and is mandatory where behaviour cannot be so bounded.

\gnote{SUGGEST EQUATION: formalise the admission state as a per-type predicate, $\mathrm{active}(E)$ over formal suability, attachment, and substantive non-emptiness, with dormancy as its negation within a type; every gate conjoins it.}

Because the type fixes the legal regime, it also fixes what may be disclosed, and this is the reason the taxonomy must be exact. Identification proves that an accountability path exists, that an act was a machine's, attached to a natural person, acting for a juridical entity, without naming any of them until an act requires it; and when names are required, the type selects the regime under which each is released. A natural entity is disclosed under the data-protection regime of its governing jurisdiction, as a data subject whose particulars are minimised and given up only on need. In EU law, a registered juridical entity is publicly disclosed through company-register machinery, while the GDPR protects natural persons and leaves legal persons outside its data-subject scope.\cite{eu2017companylaw}\cite[recital~14]{eu2016gdpr} The natural persons behind a juridical entity keep their protection when the entity is read: the Court of Justice of the European Union invalidated general public access to beneficial-ownership registers on that ground, preserving legal-entity transparency while the persons within it retain data-protection interests.\cite{cjeu2022luxembourgbusinessregisters} A machine entity discloses its attachment path, which resolves to a natural or juridical entity and is then governed by that entity's regime. A sole proprietor confirms the rule rather than breaking it: a natural person trading under a registered business name is disclosed as a natural entity with a disclosure obligation added by the act of registering, not reclassified as a juridical one, because disclosure tracks type, not activity.

A worked failure shows the path the taxonomy exists to reconstruct. Suppose a machine mishandles and loses information it was processing. The system resolves the machine's identity and reads its attachment as it stood at the time of the act, which names a natural person. That person acted as an organ of a juridical entity within a valid mandate, so the trace first imputes the consequence to the juridical entity that determined the processing purpose and means, while controller obligations, processor duties, compensation, and any internal or personal recourse are then evaluated under the applicable data-protection and organisational law.\cite[arts.~4(7), 24, 28, and 82]{eu2016gdpr} Had the machine been third-party software, the attachment would instead have named its vendor, a further juridical entity, without introducing a new type. The example is an assurance trace, not a full liability judgment: the path is replayed from the point-in-time, liability-allocated record of the act under the law in force when the act occurred.

\subsection{Jurisdictional anchoring and temporal attribution}
The multi-jurisdictional design treats jurisdictional anchoring as a time-indexed attribute of the entity's assurance state, selected for an actor when an act is evaluated. A natural entity may live in Austria at one time, move to Canada later, retain Austrian citizenship, hold a United States passport, add a Canadian residence artifact, and continue to use the same legal identity-assurance state. The state records parallel jurisdictional artifacts and the valid time of each artifact, while the requested act selects the anchor and legal predicates relevant to its evaluation.

Stated minimally, let $A_E(t)$ be the set of jurisdictional artifacts valid for entity $E$ at act time $t$. The active anchor for act $a$ is the selected member
\begin{equation}\label{eq:anchor}
  \alpha(E,a,t) \;\in\; A_E(t),
\end{equation}
chosen by the legal context of the act, the entity type, and the gate that later consumes the state. Other memberships remain in the assurance state as corroborating, historical, or capability-specific artifacts, while $\alpha(E,a,t)$ fixes the legal frame under which the act is read.

This separates stored facts from jurisdiction-relative predicates. A date of birth, address artifact, citizenship, residence status, tax status, register entry, or mandate can be stored as a fact with its own valid time. Whether that fact makes the actor of age, resident, authorised, eligible, regulated, protected, or disqualified is a predicate evaluated under the active anchor and the governing law of the requested act. The same fact can satisfy one gate in one jurisdiction and fail another elsewhere without changing the fact itself.

Account-bound systems lose this distinction. If an account is permanently bound to the jurisdiction in which it was created, later life events become structurally misrepresented: a person who moves may be treated as if the original country still governs all capability decisions, or may be forced to abandon the account and create a new one. Legal identity assurance separates entity continuity from jurisdictional state. The entity remains the same assurance-state bearer across time, and the actor remains the accountable point in each act, while the active jurisdictional anchor, residence, tax status, address, regulatory eligibility, and capability predicates can change and remain auditable.

Imputation also carries a clock. An act is evaluated against the legal and policy state in force at the time relevant to that act, while later audit may occur under a different record state and a different legal environment. Bitemporal records, valid time held alongside transaction or record time, keep that distinction stable across a life that moves between jurisdictions.\cite{snodgrass1992temporaldatabaseglossary} Later reliance records consume this structure: they can replay which artifacts were valid, which anchor was active, and which legal predicates governed the act at the time it occurred.

\subsection{Requirements for multi-jurisdictional legal identity assurance}
The preceding derivation supplies the requirements against which the artifact is later evaluated. They are stated only after the taxonomy and jurisdictional anchor because two of them need those constructs: machine action can be assessed only once M has been defined as a technical actor with borrowed legal accountability, and jurisdictional continuity can be assessed only once the anchor function has been introduced. The requirements are design constraints. A candidate assurance architecture satisfies them when its data structures, gates, and reliance records can enforce the constraint without relying on institutional habit alone. The checklist in \cref{tab:requirements} translates that derivation into six testable constraints, linking each requirement to its legal or standards pressure and to the construct that discharges it.

\begin{table}[!htbp]
\centering
\footnotesize
\setlength{\tabcolsep}{3.5pt}
\renewcommand{\arraystretch}{1.18}
\begin{tabularx}{\textwidth}{>{\raggedright\arraybackslash}p{0.11\textwidth} >{\raggedright\arraybackslash}p{0.30\textwidth} >{\raggedright\arraybackslash}p{0.29\textwidth} >{\raggedright\arraybackslash}X}
\toprule
Req. & Constraint & Legal or standards pressure & Discharged by \\
\midrule
R1 & Identity demand must be proportional to the requested act and its consequence weight. Low-consequence participation must be available at the lower floor. & Data minimisation requires processing limited to what is necessary; NIST requires a relying party to determine whether proofing is needed before selecting an assurance level.\cite[art.~5(1)(c)]{eu2016gdpr}\cite[secs.~3.3.2--3.3.3]{nist80063r4} & \Cref{eq:assurance}, \cref{eq:demand}, and capability gates that request only the cells each act consumes. \\
R2 & Assurance state must remain separate from capability permission. The same state may pass one gate, fail another, and remain auditable. & eIDAS and NIST define assurance baselines, while service authorization remains a relying-party decision.\cite{eu2014eidas,eu2015eidaslevels,nist80063r4} & The assurance state, demand set, and gate pass relation, rather than a stored verified/unverified account flag. \\
R3 & The model must support N, J, and M without making technical action collapse into legal personhood. Consequential machine action must attach to N, J+N, J+N+M, or a legally sufficient J+M path. & AI and product-liability instruments allocate duties and liability to legal actors around systems; AML beneficial-ownership rules ground juridical accountability in natural persons.\cite{eu2024aiact,eu2024pld,ep2020ailiability}\cite[arts.~51--63]{eu2024amlr} & The entity taxonomy, attachment rule, inheritance bound, and constellation grammar. \\
R4 & Disclosure must be granular enough that a gate can request a named assertion class without receiving a packed identity record. & GDPR data minimisation and data protection by default require purpose-scoped processing; the European Digital Identity Framework supports selective disclosure through the wallet architecture.\cite[arts.~5(1)(c) and 25(2)]{eu2016gdpr}\cite[arts.~5a and 5b(9)]{eu2024eidas2} & Assertion rows, optionality, cardinality, artifact multiplicity, and selective presentations. \\
R5 & The assurance state must preserve jurisdictional continuity across moves, parallel statuses, and later audits. Facts and jurisdiction-relative predicates must remain separable. & Cross-border electronic-identity frameworks support reusable identity means, and credential standards support holder-presented credentials across verifier contexts, while legal evaluation remains tied to the governing frame of the act.\cite{eu2024eidas2,w3c2025vcdm} & Jurisdictional artifacts, the active-anchor function $\alpha(E,a,t)$ in \cref{eq:anchor}, and bitemporal valid-time and record-time storage. \\
R6 & Reuse must allocate responsibility for reliance instead of erasing it. A relying party may consume prior confirmation, but its legal responsibility and the relied-on snapshot must remain visible. & AMLD4 permits third-party reliance while retaining ultimate responsibility with the obliged entity; FATF Recommendation~17 and the AML Regulation preserve a risk-based due-diligence structure.\cite[art.~25]{eu2015amld4}\cite[rec.~17]{fatf2025recommendations}\cite[arts.~19, 20, 33, and 34]{eu2024amlr} & Confirmation events, gate-pass records, reliance snapshots, requirement versions, parties, and liability boundaries. \\
\bottomrule
\end{tabularx}
\caption{Design requirements for the proposed legal identity-assurance artifact. The requirements become the criteria for the comparative evaluation.}
\label{tab:requirements}
\end{table}

The table also fixes the boundary between the general assurance layer and sectoral regulation. Anti-money-laundering simplified, standard, and enhanced due diligence are treated as one upper mapping over the architecture. They supply concrete gate predicates where the requested capability falls inside the AML domain, while the general artifact remains broader: it defines the entity, the evidence cells, the jurisdictional anchor, the gate, and the reliance record that other legal mappings can also consume.

\subsection{Assertion granularity and data minimisation}

The assertion axis is built first, and it is built from the demand side. The demand set of \cref{eq:demand} presupposes that an act can name the facts it requires, $R(a)$, and receive those and nothing else; the axis must be granular enough that every legally distinguishable demand is separately addressable. The unit of the axis is the \emph{assertion class}: one kind of identity content an entity can reveal, a date of birth, a contact channel, a civil name, a government identity document. A class is one row of the type's catalogue, and the catalogue is the complete ordered list of classes the type admits.

The legal driver of this granularity is data minimisation: personal data must be adequate, relevant, and limited to what is necessary for the purpose of the processing.\cite[art.~5(1)(c)]{eu2016gdpr} The duty is stated against the collector, but it is the data structure that decides whether the duty is enforceable. A packed identity record, the single \enquote{identity verified} object carrying name, birth date, address, document scans, and biometrics as one block, makes minimisation unenforceable in practice: a gate that needs one item receives the block, because the block is the only addressable object. The packed instance also sets the incentive. Where the marginal cost of asking is zero and the record arrives whole, every service collects the maximum, and the regulator is left auditing intentions rather than structures. Row granularity inverts this. When each class is a separately disclosable object, a gate can request a class only by naming it, an unrequested class is structurally undisclosed rather than withheld by policy, and minimisation stops being a promise and becomes a property of the address space.

What belongs on the axis is bounded by a second cut: the boundary between legally required and functionally needed information runs along the actor's role in the interaction, not along datum type. The same e-mail address is assurance-relevant where it is the confirmed contact channel a recovery or service-of-process duty relies on, and purely functional where it is a newsletter field; the same image is a confirmed likeness in a verification ceremony and ordinary content in a gallery upload. Functional data stays service-local and outside the assurance state; the catalogue carries only content that some gate predicate or legal mapping can demand.

The rows are ordered by willingness to reveal, the disclosure cost the entity bears, not by where the evidence lives or how strong it can be made; rows may shift within a small band as practice teaches, but rows never merge, because a merged row is an address the demand set can no longer form. Two row properties complete the machinery. The first is \emph{optionality}: exactly one row of every catalogue is mandatory, the existence-status row introduced with the taxonomy, row zero of every type, and every other row is voluntary, empty being a legitimate state of a voluntary row. An entity holding only its status row is a welcome participant that can do nothing requiring trust: the floor of the model is presence without suspicion, the structural opposite of the flat design's suspicion at the door. The second is \emph{cardinality}: a class admits either one slot or an open set of instances. A civil name and a date of birth are single-slot; contact channels, addresses, government documents, and payment routes are open sets, because a person legitimately holds several of each.

An open-set class holding several independently confirmed instances exhibits \emph{artifact multiplicity}, and the model's reading of it is load-bearing: more artifacts at the same class raise corroboration, never the level. Three passports from three issuing states remain one class, the government identity document; what grows is the corroboration of that class, since three issuers share no failure mode, and the willingness to provide further instances is itself a signal. Writing $\mathcal{C}_E(r)$ for the set of confirmations standing behind the artifacts an entity $E$ holds at class $r$, the \emph{confirmation index}
\begin{equation}\label{eq:confindex}
  c_E(r) \;=\; \bigl|\, \mathcal{C}_E(r) / \!\sim \,\bigr|
\end{equation}
counts confirmations up to the equivalence $\sim$ of shared failure modes, confirmations with a common issuer, channel, or custodian collapsing into one. The index is corroboration, never rank: it strengthens reliance on a class without moving the entity anywhere on any scale. This three-passport property is what every single-scale design mis-reads as a level change, and it returns as a named fold in \cref{sec:coordinate}.

\subsection{Source-of-information scale}
\begin{figure}[!htbp]
\centering
\begin{tikzpicture}[
  rootbox/.style={draw, dashed, rounded corners=3pt, minimum height=0.85cm, minimum width=4.4cm,
                  align=center, font=\scriptsize},
  groupbox/.style={draw, dashed, rounded corners=3pt, align=center,
                  font=\scriptsize, text width=2.6cm, minimum height=0.85cm, inner sep=2.5pt},
  headbox/.style={draw, rounded corners=3pt, fill=black!4, align=center,
                  font=\scriptsize, text width=2.05cm, minimum height=1.0cm, inner sep=2.5pt},
  notebox/.style={draw, dashed, rounded corners=2pt, align=center,
                  font=\fontsize{5}{6}\selectfont, text width=2.05cm, minimum height=0.75cm, inner sep=2.5pt},
  conn/.style={semithick, dashed},
]
\node[rootbox] (root) at (0,0) {Source of Information\\[-0.25em]{\fontsize{6}{7}\selectfont who provided or confirmed the information}};
\node[groupbox] (pnp) at (-7.35,-1.8)  {Provided by a natural person};
\node[groupbox] (vm)  at (-3.675,-1.8) {Verified by machine};
\node[groupbox] (vnp) at (3.675,-1.8)  {Verified by natural person};
\node[groupbox] (ta) at (1.225,-3.05) {Trained verifier};
	\node[groupbox] (no) at (6.125,-3.05) {Public or qualified};
\node[headbox] (s1) at (-7.35,-4.45) {Self\\S1};
\node[headbox] (s2) at (-4.9,-4.45)  {Machine\\S2};
\node[headbox] (s3) at (-2.45,-4.45) {Register API\\S3};
\node[headbox] (s4) at (0,-4.45)     {Digital\\S4};
\node[headbox] (s5) at (2.45,-4.45)  {In person\\S5};
\node[headbox] (s6) at (4.9,-4.45)   {Digital\\S6};
\node[headbox] (s7) at (7.35,-4.45)  {In person\\S7};
\node[notebox] (c1) at (-7.35,-5.85) {provision only; the floor};
\node[notebox] (c2) at (-4.9,-5.85)  {automated machine check};
\node[notebox] (c3) at (-2.45,-5.85) {record queried, compared};
\node[notebox] (c4) at (0,-5.85)     {verifier, remote channel};
\node[notebox] (c5) at (2.45,-5.85)  {verifier, co-present};
	\node[notebox] (c6) at (4.9,-5.85)   {public or qualified, remote};
	\node[notebox] (c7) at (7.35,-5.85)  {public or qualified, co-present};
\draw[conn] (root.south) -- (0,-0.85);
\draw[conn] (-7.35,-0.85) -- (3.675,-0.85);
\draw[conn] (-7.35,-0.85) -- (pnp.north);
\draw[conn] (-3.675,-0.85) -- (vm.north);
\draw[conn] (3.675,-0.85) -- (vnp.north);
\draw[conn] (pnp.south) -- (s1.north);
\draw[conn] (vm.south) -- (-3.675,-3.5);
\draw[conn] (-4.9,-3.5) -- (-2.45,-3.5);
\draw[conn] (-4.9,-3.5) -- (s2.north);
\draw[conn] (-2.45,-3.5) -- (s3.north);
\draw[conn] (vnp.south) -- (3.675,-2.45);
\draw[conn] (1.225,-2.45) -- (6.125,-2.45);
\draw[conn] (1.225,-2.45) -- (ta.north);
\draw[conn] (6.125,-2.45) -- (no.north);
\draw[conn] (ta.south) -- (1.225,-3.7);
\draw[conn] (0,-3.7) -- (2.45,-3.7);
\draw[conn] (0,-3.7) -- (s4.north);
\draw[conn] (2.45,-3.7) -- (s5.north);
\draw[conn] (no.south) -- (6.125,-3.7);
\draw[conn] (4.9,-3.7) -- (7.35,-3.7);
\draw[conn] (4.9,-3.7) -- (s6.north);
\draw[conn] (7.35,-3.7) -- (s7.north);
\foreach \i in {1,...,7} {
  \draw[conn] (s\i.south) -- (c\i.north);
}
\node[font=\fontsize{6}{7}\selectfont] at (0,-6.8) {accountable weight increases S1 $\rightarrow$ S7; within the human and public/qualified tiers, co-presence outranks the remote channel};
\end{tikzpicture}

\smallskip
\gnote{Leaf labels, state 2026-06-10: S3 RULED, Register API (wide reading: the system of record per row, state, bank, carrier; replaces eGov and authoritative source). Group label ``provided by a natural person'' RULED to stand, because a juridical entity cannot act by itself, so provision always grounds in a natural will (J+N minimum; J+M deferred); mirror of the behaved-through cut and of constellation pattern E being restricted. Still open to DECIDE: Physical vs in person (S5/S7), Trained Agent vs trained verifier, Legal Notary vs notary; Entity for S1 was dropped (collides with the root of \cref{fig:entity-taxonomy}).}
\caption{The source-of-information scale, the second axis. The classes of the scale are exactly the seven grey leaves, S1--S7; the outlined nodes above them are explanatory grouping, not a taxonomy, and the dashed boxes beneath are glosses. A statement is either provided, by the subject itself, the self (S1), the floor of the scale because an entity cannot confirm itself, or verified: by machine, an automated check (S2) or a register API, the system of record for the fact queried and compared (S3), or by a natural person, where the trained-verifier tier (digital S4, in person S5) and the public or qualified confirmation tier (digital S6, in person S7) each split into a digital and an in-person class, because co-presence removes the remote-channel attack surface. Where the governing framework recognises the confirming authority, the public or qualified tier outranks a trained human verifier, so the in-person human class (S5) sits below the digital public or qualified class (S6), and the in-person public or qualified class (S7) is the strongest source class. The two sub-axes under each class are the source itself and the strength of its binding to the subject; what an individual confirmation is worth at a particular gate is derived by the relying party from the cell's position, binding, and freshness, never stored as a score.}
\label{fig:source-scale}
\end{figure}
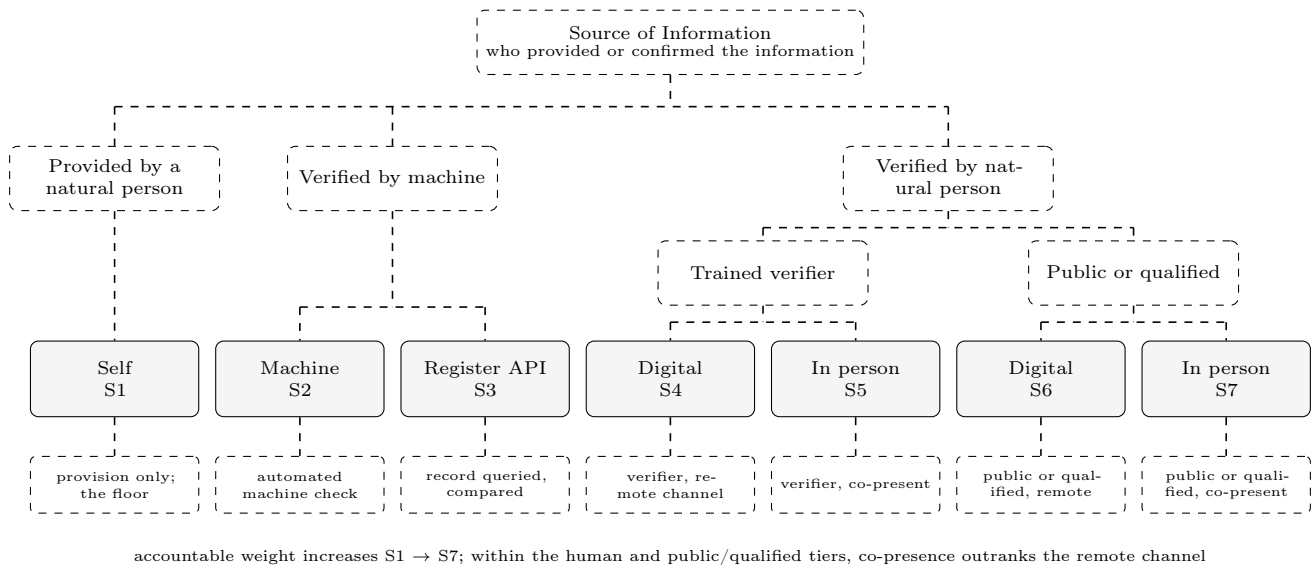

The source-of-information axis is the second axis on which a cell is read, and it is the mirror of the cut that types entities. Where the entity axis asks who can be held for an act, this axis asks who can be held for a confirmation, how much independent and accountable weight stands behind a disclosed assertion. It does not measure whether the subject is trustworthy; we reserve the word trust for the gate-side quantities derived from reliance. It states only the evidentiary strength of the assertion as presented, and it does so on one homogeneous scale, drawn in \cref{fig:source-scale}, shared by all three entity types; what differs by type is which confirmers can occupy each class for a given item, developed with the per-type catalogues.

The scale has the self as its floor, because an entity cannot confirm itself. A self-declared assertion is provision, not verification: it records that the subject said so, which may be true or false, presumed unreliable by design, and that presumption is the premise the whole system answers. Every step above the floor adds a confirmer that is independent of the subject and that answers, in some degree, for the confirmation being wrong, so the scale rises by the accountability of the confirmer rather than by the mechanism it uses. A machine confirmation (S2) is independent of the subject's say-so, an automated check, a cryptographic signature, a liveness test, but carries only the accountability of whoever deployed it, so its strength tracks what stands behind the check, not the fact that a machine performed it.\cite{nist80063a4,enisa2024ridpgoodpractices} A register-API confirmation (S3) queries the system of record for the fact and compares: which register that is varies with the row, a population register for a birth date or civil name, the commercial register for a company, the bank for its own account relationship, the carrier for a number, in each case an institution under a legal or regulated-record duty to hold that record correctly.\cite{eu2017companylaw,fatf2020digitalidentity} A human confirmation adds an identified, trained verifier who can be held to the check; a notarial or qualified public confirmation adds a legally empowered officer or trust-service proofing context whose attestation carries higher accountable weight in the relevant jurisdiction or scheme.\cite{hcch1961apostille,etsi2025ts119461,uncitral2022mlit}

The human and public/qualified tiers each split into a digital and an in-person class, and the split is categorical rather than presentational. A remote ceremony and a physically co-present one differ in attack surface: co-presence removes the remote channel, and it changes what the verifier can do, a document turned in the hand rather than presented to a camera, a person attending in body rather than in pixels.\cite{goffman1967interaction,enisa2024ridpgoodpractices,etsi2025ts119461} Every class on the scale is forgeable in principle, that is the premise the system rests on; a residual attack against a class, a deepfake against a remote ceremony, is priced into the evidential weight a relying party derives from the cell, never used to collapse the class, because the cost, the legal exposure, and the per-account effort of mounting it are part of what the class is worth. The ordering that results is S4 human digital, S5 human in person, S6 public or qualified digital, S7 public or qualified in person: within each tier co-presence outranks the remote channel, while the public officer's or qualified confirmer's legal accountability can outrank the trained-human tier, so the in-person human verification still sits below the digital public or qualified class where the governing jurisdiction or scheme recognises that attestation.

Self (S1), machine (S2), register API (S3), human digital (S4), human in person (S5), public or qualified digital (S6), and public or qualified in person (S7) are named classes of this scale, ordered by the accountable weight standing behind the confirmation, not by the mechanism it uses. A gate may demand either a floor, at least S3, or a specific source kind where the mechanism matters. Two distinctions sit under the single coordinate. The first is between a source of truth and a source of confirmation: an authoritative register is the canonical origin of a fact, whereas a public or qualified confirmer verifies a presented claim against it; the model treats both as accountable weight behind the cell and privileges neither form. The second is binding: a confirmation is only as strong as its attachment to this subject and this instance, so each class composes the accountability of the source with the strength of the binding between the confirmed fact and the entity at the gate. A register entry that cannot be bound to the presenter confirms a fact about someone, not about the actor.

Three rules attach to the register-API class, and the first reaches beyond it. A confirmation event seals exactly the row-facts the confirming act examined, never an index range: one register call that validates a birth date, a civil name, a likeness, and a document number against the record fills the S3 cell of each of those rows in a single event, just as an in-person public or qualified act covers the facts the presented document carries. Cells filled by one event are correlated, one call, one failure mode, so they never count as independent corroboration in the confirmation index of \cref{eq:confindex}, and they retract together if the event itself proves spoofed. Second, the class has two forms. A data pull queries the register about someone and confirms the fact while binding no presenter; a holder-triggered connect, a wallet confirmation or an electronic-identity PIN, has the subject reach into the register from inside its own control, supplying source and sole-control binding in one act. Both are one class, the binding recorded on the confirmation event. Where the record holds a reference likeness, the register supplies the fact while the live match of the presenter against it remains a machine act (S2); the two cells stay apart. Third, a confirmation is an act, and a party that has not acted in the system has confirmed nothing in it. An external verification provider enters in one of two ways: as an issuer, transmitting confirmation events that inherit the class of the underlying check it can document and that are worth no more than their weakest verifiable link, or as the trained human verifier of S4 and S5, when it actually confirms the person rather than forwarding facts about one.

Two rules follow from reading strength as accountable weight. A confirmation assembled from several links is only as strong as its weakest required link, the same weakest-link rule that governs the inheritance bound, so a notarised document resting on an unverified underlying claim is no stronger than that claim. And confirmation cascades: a higher source may attest the result of a lower one, a public or qualified confirmer attesting that a human verified a machine check, which is what lets the scale telescope a chain of confirmations into a single accountable point.

\gnote{Open point i RULED in principle (2026-06-10): the coverage rule stated earlier in this subsection, an event seals exactly what it examined, never an index range; it covers both the register call and the notary master seal. SUGGEST EQUATION: the formal derived-cell semantics (event pointer, retraction, no independence credit) still to write for the event model.}

The scale is one and the same for all three entity types; only the confirmers that can occupy each class change with the type. The register of record at S3 is a population register for a natural entity and the commercial register for a juridical one; the public or qualified confirmer at S7 attests a person's deed or a company's instrument in physical presence where that function is available. A machine entity is confirmed through the confirmers its descriptors admit on the same scale: a cryptographic attestation of the running artifact is a machine confirmation (S2), a conformity assessment or notified-body attestation a register confirmation (S3), an independent audit an accountable human verification (S4, S5), and a notarised, public-officer, or qualified mandate confirmation a high-accountability public or qualified confirmation (S6, S7). On any catalogue, where no source of a given class exists for an item, the cell is inapplicable rather than failed: a biometric pattern has no authoritative register, an unregistered association no commercial-register entry, a stateless person no national identity source. The model records this as a first-class empty value, distinct from an item left unconfirmed, so that an entity reaches assurance through the classes available to it rather than being penalised for the absence of a class it could never use. This is what lets the graded scale admit participants that a flat maximum-verification bar would exclude, and it is the rule forecast in the related work: assurance stays reachable through the strongest recognised non-sovereign confirmation channel available for the relevant fact and jurisdiction, so a sovereign scheme accelerates the state without ever being a precondition for it.

\gnote{DECIDE: name the rule in the last sentence (working name: the equity rule); the Related Work forward reference points here. Open point ii RULED (2026-06-10): S3 is the register API, wide reading, the system of record per row (state, bank, carrier); two forms, data pull (no binding) and holder-connect (sole-control binding), one class with binding on the event; chip read stays S2; external KYC providers are issuers or S4/S5 humans, never a class. Remaining open: whether the two forms get a sub-notation.}

\subsection{Per-type assertion catalogues}
%

The natural catalogue is the development root, and its rows are derived by walking the willingness ordering from the floor upward, each row admitted because some gate predicate or legal mapping can demand it and because it is separately disclosable. N0, the status row, is the mandatory existence anchor: it records that the entity exists and whether it is alive or deceased, the root fact every other row presupposes, and the only row a bare participant must fill. N1, the date of birth, is the cheapest legally consequential fact, since age predicates are the most common statutory gate, and the fact, once confirmed, does not expire. N2, contact channels, an open set, carries reachability, the operational minimum of participation, and is confirmable by control proof. N3, the civil name, single-slot, is the coordinating identifier the deception-free limit preserved: it tells parties apart and holds until a legal event changes it. N4, addresses together with the jurisdictional anchor, an open set, ties the entity to the legal frames of the jurisdictional-anchoring subsection above. N5, the personal likeness, an open set, is the binding substrate of verification ceremonies and the most freshness-sensitive row. N6, government identity documents, an open set, holds the composite artifacts a state issues; each document is one instance, separately confirmed, and the claims a document carries, a birth date, a name, remain claims of their own rows. N7, payment details, and N8, payment interfaces, are distinct open sets because their confirmers and failure modes differ: a bank relationship is confirmed by the regulated institution holding it, a payment-interface connection by control proof against the provider. N9 and N10, the voice and face patterns, close the catalogue: biometric patterns sit last because their disclosure cost is highest and least reversible, and they exhibit the inapplicable cell, since no register of record holds them. \cref{tab:catalogue-natural} renders the catalogue against the source scale.

\begin{table}[!htbp]
\caption{Assertion catalogue of the natural type: eleven classes N0--N10, ordered by willingness to reveal, with optionality (mand mandatory, vol voluntary) and cardinality (1 one slot, n an open set of instances). The source-availability note states ordinary confirmers; the binary S1--S7 cells are entity-specific and are filled by confirmation events.}
\label{tab:catalogue-natural}
\centering
\footnotesize
\begin{tabularx}{\textwidth}{@{}>{\raggedright\arraybackslash}p{3.2cm}p{0.8cm}p{0.8cm}X@{}}
\toprule
Class (content) & Opt & Card & Source-availability note\\
\midrule
N0 Status & mand & 1 & civil register, population register, death register, public authority, or self-declared floor\\
N1 Date of birth & vol & 1 & civil register, identity document, wallet confirmation, or public/qualified attestation\\
N2 Contact channels & vol & n & control proof, carrier or provider record, or verified contact ceremony\\
N3 Civil name & vol & 1 & civil register, identity document, name-change record, or public/qualified attestation\\
N4 Addresses + jurisdiction link & vol & n & population register, residence permit, tax record, utility or bank evidence, or public/qualified attestation\\
N5 Personal likeness & vol & n & live capture, document-photo comparison, biometric ceremony, or public/qualified witnessing\\
N6 Government ID & vol & n & issuing authority, chip or wallet confirmation, register query, or public/qualified attestation\\
N7 Payment details & vol & n & regulated bank record, account confirmation, payment-institution record, or mandate proof\\
N8 Payment API & vol & n & provider control proof, token binding, payment-interface confirmation, or bank-mediated connection\\
N9 Voice pattern & vol & n & live capture, biometric ceremony, device-bound sample, or public/qualified witnessing\\
N10 Face pattern & vol & n & live capture, biometric ceremony, device-bound sample, or public/qualified witnessing\\
\bottomrule
\end{tabularx}
\end{table}

The juridical catalogue follows the same construction over the items the taxonomy already identified. Its row zero, J0, records whether the entity is active, dissolved, suspended, or in insolvency, mandatory like every existence-status row. J1 records the legal name under which the entity is addressed. J2 records the juridical form and governing law, because company, association, foundation, partnership, public body, and trustee-imputed arrangement differ by legal form rather than by type. J3 records the founding instrument or constitutive act. J4 records the register, public-law instrument, or recognised non-register basis that anchors the entity. J5 records organs, offices, representatives, trustees, or other authority-bearing positions. J6 records purpose, capacity, and activity limits. J7 records the accountable natural persons, beneficial owners, controllers, partners, trustees, or fiduciary officeholders behind the entity.\cite{eu2017companylaw,fatf2023beneficialownershiplegalpersons,fatf2024beneficialownershiplegalarrangements} \Cref{tab:catalogue-juridical} turns these legal-addressing items into row addresses, so a gate can ask for organisational status, authority, capacity, or control-path evidence without treating the juridical entity as if it acted by itself.

\begin{table}[!htbp]
\caption{Assertion catalogue of the juridical type: eight classes J0--J7. The rows cross the same source scale S1--S7 as the natural catalogue; the source-availability note states the ordinary high-weight confirmer without excluding other documented confirmations.}
\label{tab:catalogue-juridical}
\centering
\footnotesize
\begin{tabularx}{\textwidth}{@{}>{\raggedright\arraybackslash}p{3.1cm}p{0.8cm}p{0.8cm}X@{}}
\toprule
Class (content) & Opt & Card & Source-availability note\\
\midrule
J0 Status & mand & 1 & register, competent authority, insolvency record, or fiduciary record\\
J1 Legal name & vol & 1 & register, founding instrument, public-law instrument, or qualified attestation\\
J2 Form and governing law & vol & 1 & register parameter or legal instrument naming the form and legal system\\
J3 Founding instrument & vol & n & deed, articles, statute, trust instrument, or public-law act\\
J4 Register or public anchor & vol & n & commercial register, supervisory register, LEI/vLEI, public act, or recognised non-register basis\\
J5 Organs and representatives & vol & n & register entry, mandate, role credential, board record, trustee office, or public appointment\\
J6 Purpose and capacity & vol & n & constitution, licence, register object, public mandate, or regulatory authorisation\\
J7 Accountable persons and control path & vol & n & beneficial-ownership, partner, trustee, fiduciary, role, or control evidence\\
\bottomrule
\end{tabularx}
\end{table}

The juridical table fixes row addresses, but it does not make a juridical actor act alone. The composite remains outside the row list. A gate over J reads the entity's J-cells, the relevant natural persons' N-grids, and the delegation or office relation that connects them under the inheritance bound of \cref{eq:bound}. The register parameter also remains a parameter, not a type split: where a commercial register exists, it is the ordinary S3 source for rows such as J0, J1, J2, J4, and J5; where the juridical form is a public body, foundation, partnership, association, or trust-like legal arrangement, the source is the competent public instrument, supervisory register, trustee record, or fiduciary proof recognised by the governing law. A cell whose source class cannot exist for that form is marked inapplicable rather than failed.

The machine catalogue is shorter because the machine has no catalogue of life-facts. M0 records existence, admission status, and attachment path: active, abandoned, blacklisted, suspended, or retired, together with the machine-containing accountability path, N+M, J+N+M, or the restricted J+M path where the governing framework supplies direct juridical responsibility for the machine's acts. M1, M2, and M3 are capability rows, preserving the capability classes already introduced in the taxonomy. M1 is a deterministic actuator whose decision path contains no model inference. M2 is a model-driven actuator acting inside a human-authored mandate. M3 is a reserved goal-setting class for an actuator that sets goals beyond a human-authored mandate. Each machine row carries two descriptors as cell metadata, not as additional rows: a trust vector, provenance, artifact integrity, audit status, explainability, and data-sovereignty claims; and a control vector, immutability, upgrade authority, administrative custody, live oversight, live recovery, and recovery surface. \Cref{tab:catalogue-machine} makes that restriction visible by putting the attachment path in M0 and treating the remaining rows as capability evidence that gates may accept only through an admissible constellation.

\begin{table}[!htbp]
\caption{Assertion catalogue of the machine type: four classes M0--M3. M1--M3 are capability rows; trust and control are descriptors attached to the machine row and its confirmation events.}
\label{tab:catalogue-machine}
\centering
\footnotesize
\begin{tabularx}{\textwidth}{@{}>{\raggedright\arraybackslash}p{3.0cm}p{0.8cm}p{0.8cm}X@{}}
\toprule
Class (content) & Opt & Card & Source-availability note\\
\midrule
M0 Status and attachment path & mand & 1 & artifact attestation, register or conformity record, audit, mandate, public-officer attestation, or role proof\\
M1 Deterministic actuator & vol & n & code hash, deployment attestation, reproducible build, bytecode proof, audit, or notarised mandate\\
M2 Model-driven mandated actuator & vol & n & model provenance, version record, risk classification, conformity assessment, oversight record, audit, or mandate\\
M3 Reserved goal-setting actuator & vol & n & future legal and technical confirmation only; active gates require explicit law, control, and recovery predicates\\
\bottomrule
\end{tabularx}
\end{table}

This row assignment keeps the machine definition aligned with the legal point made earlier. M0 is mandatory because an active machine must expose where responsibility attaches before its output can matter legally. M1 can be bounded through pre-admission analysis of fixed behaviour. M2 needs mandate, oversight, and recovery evidence because its behaviour is bounded through an operating envelope rather than through fixed code. M3 remains reserved in the catalogue so that future goal-setting systems have an address, while present gates can decline active capability until the governing framework supplies accountability, control, and recovery conditions. Confirmation runs on the shared source scale through the confirmers the row admits: artifact checks at S2, registers or conformity bodies at S3, audits at S4 and S5, and notarised, public-officer, or qualified mandate confirmations at S6 and S7.\cite{eu2024aiact,eu2024pld,ep2020ailiability,uncitral2024automatedcontracting}

\subsection{Legal identity-assurance coordinate and event model}
\label{sec:coordinate}
%
%
The first axis is not a moral or behavioural score. It is a disclosed identity-scope axis. For a natural actor, N0 through N10 represents the scope of identity assertions that are available for, required by, and disclosed into a concrete capability context. It combines four design facts: what the gate requires, what the applicable law or regulatory mapping requires, what the actor has provided into the holder-carried assurance state, and what the actor signs off for disclosure in the particular presentation. Provision and disclosure remain distinct: an actor may hold an artifact in the wallet without disclosing it to a relying service.

The second axis is the source-of-information axis. It asks how reliable the disclosed assertion could be, given who or what stands behind it. A self-provided date of birth at S1 may be true or false; the model records it as self-provenanced rather than externally confirmed. A public or qualified confirmation of date of birth at S7 carries a different evidentiary status because a legally accountable confirmation source stands behind it. The source axis does not claim that the actor is trustworthy. It states the evidentiary strength of the disclosed assertion.

The combined coordinate should be read as low or high disclosed assurance, not as good or bad identity. N1S1 is a legitimate state: the actor may be newly enrolled, may have no reason to reveal more, may hold stronger artifacts without disclosing them, or may indeed be fake. The system response is capability containment rather than punishment. A low-disclosed-assurance actor can exist in the system while being unable to exercise capabilities whose legal or operational consequences require stronger attribution, such as unsolicited connection, public posting, business contracting, payment, or asset movement.

Cross-jurisdictional artifacts are instances of this model, not exceptions to it. A natural actor may hold an EUDI personal identification data artifact, a United States passport, a United States state identity document, an Austrian address artifact, a Canadian residence artifact, bank-control evidence, or other jurisdictional N instances. The legal identity-assurance state stores these artifacts and their temporal validity, while a capability gate consumes only the signed package needed for the requested action under the relevant jurisdictional and regulatory mapping.

The two axes compose into a single coordinate per disclosed assertion, the cell, and an entity's assurance state is the set of cells it has filled, one assertion item confirmed at one source. Because a strong cell cannot substitute for a missing one, the state is a set, not a summed score.


Stated formally, an entity $E$ of type $T$ carries the catalogue $R_T$, and its assurance state is the binary grid
\begin{equation}\label{eq:grid}
  G_E \;\subseteq\; R_T \times \{\mathrm{S1},\dots,\mathrm{S7}\},
\end{equation}
a cell $(r,s)$ filled exactly when a confirmation by source class $s$ exists for an artifact the entity holds at class $r$. The grid is the whole state: each filled cell carries its confirmation dates, time being an attribute of the cell rather than a third axis, and the per-class confirmation index of \cref{eq:confindex} is derived from it. A \emph{coordinate} names one cell; in this paper the notation uses the prefix \texttt{EAID}, written \texttt{EAID-N6-S7}. The coordinate is a ruler and an address at once, and a single verification flag is neither. As a ruler, the set of filled coordinates reads on both axes together: across the assertion axis how much the entity has revealed, down the source axis how strongly each revealed piece is confirmed, per piece and never averaged. As an address, the grid is navigable in both directions: any artifact resolves forward to one cell, and any cell resolves backward to the exact artifacts and confirmations standing behind it. Selective disclosure, audit, and portability all rest on this two-way addressability, and it is what a flat \enquote{verified: yes} destroys: from the bit, nothing about what was checked, by whom, or to what strength can be recovered, and no single item can be located.

\begin{figure}[!htbp]
\centering
\begin{tikzpicture}[
  cell/.style={draw, minimum width=0.46cm, minimum height=0.46cm, inner sep=0pt},
  filled/.style={cell, fill=black!35},
  lvl/.style={draw, rounded corners=2pt, minimum width=2.2cm, minimum height=0.5cm,
              align=center, font=\scriptsize, fill=black!4},
  lbl/.style={font=\fontsize{6}{7}\selectfont},
]
\foreach \r in {0,...,5}
  \foreach \s in {0,...,6}
    \node[cell] at (\s*0.46,-\r*0.46) {};
\foreach \s/\r in {0/0,1/0,2/0,0/1,2/1,6/1,0/2,1/3,3/3,0/4}
  \node[filled] at (\s*0.46,-\r*0.46) {};
\node[lbl] at (1.4,0.55) {sources S1 $\rightarrow$ S7};
\node[lbl, rotate=90] at (-0.55,-1.15) {classes};
\node[lbl] at (1.4,-3.1) {the grid: every cell addressable};
\draw[->, semithick] (3.6,-1.15) -- (4.6,-1.15) node[midway, above, lbl] {fold};
\node[lvl] (l4) at (6.3,0.1)  {Level 4};
\node[lvl] (l3) at (6.3,-0.55) {Level 3};
\node[lvl] (l2) at (6.3,-1.2) {Level 2};
\node[lvl] (l1) at (6.3,-1.85) {Level 1};
\node[lbl] at (6.3,-3.1) {the ladder: one rung, no addresses};
\end{tikzpicture}
\caption{The fold. Left: the assurance state as the two-axis grid of \cref{eq:grid}, each cell separately addressable, filled cells scattered as real states are. Right: the same state folded into a per-credential ladder of levels, the form shared by flat verification and level-of-assurance designs. The fold deletes cell addressability; \cref{tab:fold-kill} lists what each deletion costs.}
\label{fig:fold}
\end{figure}
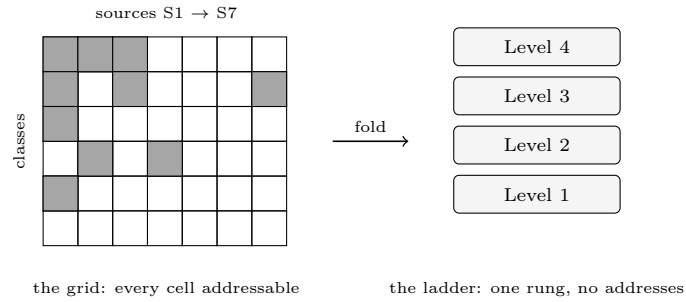

A tier, finally, is a value a gate derives from the whole filled grid against its declared threshold, never a property the grid stores and never a single cell read as a rank. This discipline is best defended by exhibiting its alternative. The assurance designs compared in the evaluation can be read as \emph{folds} of the grid in \cref{eq:grid}: projections that merge rows into bundles, merge columns into a verified bit, or read the grid's diagonal as one ordered ladder of levels, as drawn in \cref{fig:fold}.\cite{eu2015eidaslevels,nist80063r4,nist80063a4,w3c2025vcdm} A fold is not a tidier presentation of the same information. Each fold deletes the addressability of some family of cells, and with it a demand that \cref{eq:demand} could otherwise express; \cref{tab:fold-kill} names the principal folds and the capability each one amputates.

\begin{table}[!htbp]
\caption{What each fold of the grid amputates. Every simplification of the two-axis state into fewer dimensions deletes the addressability of some cells and with it a demand the gate grammar could otherwise express; the comparative evaluation below evaluates exactly these folds as designs.}
\label{tab:fold-kill}
\centering
\footnotesize
\begin{tabularx}{\textwidth}{@{}>{\raggedright\arraybackslash}p{3.2cm}>{\raggedright\arraybackslash}X>{\raggedright\arraybackslash}X@{}}
\toprule
Fold & What it deletes & Demand made inexpressible\\
\midrule
Rows merged into data blocks & per-class addressability; an unrequested class travels with its block & a gate that demands the address package at register-API source and nothing else\\
Source columns merged into one verified bit & graded confirmation strength & the same fact accepted self-declared in one jurisdiction and required at machine strength in another, with no additional data disclosed\\
Corroboration merged into rank & the distinction between level and confirmation index & three passports read as higher corroboration of the same class rather than as a level-up\\
Grid folded to enrolled / verified & the voluntary floor & a participant that exists, is welcome, and can do nothing that requires trust\\
\bottomrule
\end{tabularx}
\end{table}

The folds are also why the assurance state must be carried as the grid itself rather than as a derived summary: a summary that collapses rows, columns, or cell provenance is a fold, and each fold pays in a named capability. The comparators of the evaluation below, the flat single bar and the per-credential level ladder, are the first and the last fold institutionalised, which is what makes the comparison a test of the structure rather than of any particular parameter setting.

The grid is reusable only if its cells carry provenance as events rather than as loose document attributes. A \emph{confirmation event} is the signed record by which a source fills one or more cells of $G_E$. It records what was checked, which source class performed the check, how the checked facts bind to the entity instance, when the underlying fact was valid, when the check entered the assurance record, which status or revocation channel governs it, and which policy version the confirmer applied. Verifiable-credential and presentation standards supply the credential, proof, status, and holder-presentation substrate; the assurance event adds the paper's cell-level source and binding semantics around that substrate.\cite{w3c2025vcdm,w3c2025vcdataint,w3c2025vcbitstringstatus,openid2025openid4vp} The implementation-facing field set in \cref{tab:confirmation-event} is intentionally event-centred: it preserves enough source, binding, timing, status, and proof context to replay why a cell was accepted at a later gate.

\begin{table}[!htbp]
\caption{Confirmation event fields. The event fills one or more grid cells and preserves the source, binding, timing, status, and proof context needed for later replay.}
\label{tab:confirmation-event}
\centering
\footnotesize
\begin{tabularx}{\textwidth}{@{}p{3.1cm}XX@{}}
\toprule
Field & Meaning & Main consumer\\
\midrule
\texttt{confirmation-id} & Stable event identifier, scoped by issuer and policy version & audit trail; aggregation layer\\
\texttt{subject} & Entity identifier and entity type $T \in \{N,J,M\}$ & grid owner; gate evaluation\\
\texttt{cells} & Set of catalogue rows and source class pairs $(r,s)$ filled by this event & assurance grid; fold-kill preservation\\
\texttt{confirmer} & Issuer, register, verifier, public officer, qualified service, auditor, or machine source that performed the check & source-scale evaluation\\
\texttt{method} & Ceremony or protocol used, including remote proofing, register pull, holder-triggered connect, audit, or public attestation & source/binding interpretation\\
\texttt{binding} & Evidence that the confirmed fact belongs to this entity instance and, where relevant, to the presenting holder & replay; anti-impersonation checks\\
\texttt{fact-valid-time} & Time or interval for which the confirmed fact itself is valid & jurisdictional and temporal predicates\\
\texttt{record-time} & Time at which the confirmation entered the assurance record & bitemporal replay\\
\texttt{status-pointer} & Revocation, suspension, expiry, or withdrawal reference for the event or credential & validity checks; freshness checks\\
\texttt{policy-version} & Legal, procedural, technical, and assurance policy version applied by the confirmer & later audit under historical rules\\
\texttt{proof-package} & Signature, data-integrity proof, credential reference, and hash of disclosed evidence & integrity and non-repudiation\\
\bottomrule
\end{tabularx}
\end{table}

The event also fixes correlation. If one confirmation event fills four cells, those four cells share the same source, ceremony, policy version, and failure mode. They may satisfy four different row demands, but they cannot count as four independent confirmations. This is the formal version of the coverage rule stated in the source scale: the event seals what the confirming act examined, and it withdraws or weakens the same covered cells if its proof, status, or authority later fails. Corroboration counts across independent confirmation events, not across the number of facts a single event carried.

Read over events, the confirmation index in \cref{eq:confindex} is a quotient over event identifiers. Let $C_E(r,s)$ be the set of confirmation events that fill cell $(r,s)$, so $(r,s) \in G_E$ exactly when $C_E(r,s)$ is non-empty. Let $e \sim e'$ when two events share a confirmer, ceremony, custody path, status authority, or proof failure mode. The row-level corroboration count is then
\begin{equation}\label{eq:eventcorrelation}
  c_E(r) \;=\; \left|\left(\bigcup_s C_E(r,s)\right) / \!\sim \right|.
\end{equation}
The quotient is what prevents one ceremony that fills several cells, or several credentials resting on one compromised issuer channel, from creating artificial independence. Retraction follows the same event pointer: when an event is revoked, invalidated, or found outside its authority, every cell it filled loses that event as support.

A \emph{reliance event} is the consuming-side counterpart. It records that a relying service accepted or rejected a presentation for a concrete act under a declared gate. The confirmation event answers what stood behind the evidence; the reliance event answers who relied on it, for which capability, under which law and gate version, with which actor constellation, and with what liability boundary. Temporal database terminology separates the time at which the underlying facts were valid from the time at which the reliance record was written, making later replay possible without pretending that the later legal or data state was already true at act time.\cite{snodgrass1992temporaldatabaseglossary} \Cref{tab:reliance-event} mirrors the confirmation-event table on the consuming side, so reuse leaves a concrete record of the decision, the gate, the presented package, and the responsibility boundary.

\begin{table}[!htbp]
\caption{Reliance event fields. The event records the relying-party decision over a presentation, with the legal, temporal, and responsibility context needed for audit.}
\label{tab:reliance-event}
\centering
\footnotesize
\begin{tabularx}{\textwidth}{@{}p{3.1cm}XX@{}}
\toprule
Field & Meaning & Main consumer\\
\midrule
\texttt{reliance-id} & Stable identifier for the relying-party decision & audit trail; dispute handling\\
\texttt{relying-party} & Service, obliged entity, public body, platform, or other party consuming the presentation & liability allocation\\
\texttt{act} & Capability request, transaction, or operation to which the presentation is attached & gate selection\\
\texttt{constellation} & Submitted actor pattern, such as N, J+N, J+N+M, N+M, J+M, or M & admissibility grammar\\
\texttt{gate-version} & Published gate profile, including demand set $D(g)$, constellation family $\mathcal{K}(g)$, and governing law $\ell(g)$ & reproducible decision test\\
\texttt{presentation} & Disclosed coordinates, confirmation-event references, proof package, and selective-disclosure metadata & evidence minimisation; replay\\
\texttt{anchor} & Active jurisdictional anchor $\alpha(E,a,t)$ and legal predicates evaluated at act time & applicable-law analysis\\
\texttt{valid-time} & Time or interval over which the relied-on facts and mandates were valid & temporal predicate replay\\
\texttt{record-time} & Time at which the reliance decision was recorded & bitemporal audit\\
\texttt{freshness-result} & Per-demand age check against $\Delta$, with pass or fail reason by cell & eligibility analysis\\
\texttt{decision} & Grant, denial, escalation, manual review, or limited capability outcome & service action; recourse\\
\texttt{liability-boundary} & Party or parties retaining responsibility for reliance, including retained-responsibility third-party reliance where applicable & accountability and recovery\\
\bottomrule
\end{tabularx}
\end{table}

The reliance event is also the privacy boundary. Data-protection law asks for lawful, purpose-bound, minimal processing by design, while anti-money-laundering law permits reliance on third parties without shifting ultimate responsibility away from the obliged entity.\cite[arts.~5, 6, and 25]{eu2016gdpr}\cite{edpb2020dpbdd}\cite[arts.~25--27]{eu2015amld4}\cite[rec.~17]{fatf2025recommendations} The event structure answers both pressures: the relying party receives only the presentation its gate declared, but the record still preserves enough source, status, time, law, and responsibility metadata to show later why the capability was granted or refused.

\subsection{Capability-gate grammar and assurance separability}


This framing separates identity assurance from capability. Identity assurance describes the actor's reusable state: assertions, confirmation sources, jurisdictional anchor, freshness, revocation, and temporal validity. Capability gating describes a consuming decision: whether that state satisfies a declared rule for a concrete interaction. The same actor state can pass a low-disclosure communication gate, fail a banking gate, and later be replayed for audit under the law and policy version in force at the time.

The same premise also disciplines proxy use in gate design. Network origin, device history, contact handles, public-profile traces, or platform tenure may be operational signals in some domains, but the legally relevant state is carried by the person, juridical entity, mandate, credential, transaction, and conduct record. In a legal identity assurance model, gate predicates should attach to those legally relevant objects. Absence of non-required data is a neutral state, not a general suspicion marker. When an adverse decision is made, the record should expose the reason category, evidence class, applicable rule, and appeal or recourse path, so that accountability remains reachable without expanding routine disclosure. This aligns the architecture with lawful basis, fairness, transparency, purpose limitation, data minimisation, and data-protection-by-design duties in data-protection law.\cite[arts.~5, 6, and 25]{eu2016gdpr}\cite{edpb2020dpbdd}

Capability gates evaluate actor constellations, not only isolated actor types. A constellation records which entities participate in an operation and which relation makes the operation imputable: a natural actor acting alone, a juridical actor acting through an accountable natural actor, a juridical actor acting through a natural anchor and a machine actor, or a natural actor acting with a machine actor. The relevant question is how strongly an actor is identified and whether the requested operation may be performed by the submitted constellation at all.

The default admissibility grammar distinguishes six constellation patterns. Pattern A, N, is a natural actor acting alone and is admissible where the service accepts individual action. Pattern B, J+N, is a juridical actor acting through an accountable natural actor and is the ordinary form for organizational capability. Pattern C, J+N+M, is a juridical actor acting through a natural anchor and a machine actor, and is the default pattern for accountable organizational automation. Pattern D, N+M, is a natural actor using or delegating to a machine actor and is service-dependent. Pattern E, J+M, is technically possible but legally restricted in this artifact and becomes admissible only where the governing law and recovery mechanisms support direct juridical-machine action. Pattern F, M, is inadmissible as a standalone active actor in this artifact because machine action without an imputation anchor produces system effects without a sufficient responsibility point.\cite{eu2024aiact,eu2024pld,ep2020ailiability,uncitral2024automatedcontracting}

\gnote{SUGGEST TABLE: the admissibility grammar as a table, patterns A--F against constellation, admissibility, and rationale; this prose enumeration is the source. The J+N-minimum rule (a juridical entity acts only through a natural will; J+M deferred) is the same ruling as the S1 group label in \cref{fig:source-scale}.}

This grammar also applies to the assurance system itself. A supervisory or coordinating system-AI operated by the assurance provider is represented as J+N+M: the provider as juridical operator, an accountable natural role or office as control anchor, and the machine actor as the operational component. The same rule that applies to external actors governs internal automation, so the system demands anchored action from others and exposes its own automation through the same accountability shape.

Gates can then be declared as service-specific bundles. A high-risk asset or contract service may require a minimum legal identity-assurance coordinate such as \texttt{EAID-N8-S4} for natural participants and may accept only N, J+N, and J+N+M constellations, while excluding N+M, J+M, and M for that operation. A low-risk chat service may require only \texttt{EAID-N2-S1} and may admit N, N+M, and J+N+M, while still excluding standalone M. The gate states both the minimum assurance state and the admissible constellation family for the requested operation.

A gate is a declared bundle $g = (D(g), \mathcal{K}(g), \ell(g))$: a demand set in the sense of \cref{eq:demand}, extended per requirement with a maximum age; an admissible constellation family $\mathcal{K}(g)$; and the governing law $\ell(g)$ under which its predicates are evaluated at check time. Let $\kappa(a)$ be the submitted constellation for act $a$. A presentation passes at time $t$ when the constellation is admitted and every demanded cell is held strongly enough and recently enough,
\begin{equation}\label{eq:pass}
  \mathrm{pass}_g(E,a,t) \iff \kappa(a) \in \mathcal{K}(g) \;\wedge\;
  \forall\, (r,\, s_{\min},\, \Delta) \in D(g):\;
  \exists\, s \succeq s_{\min}:\; (r,s) \in G_E \;\wedge\; t - t_{\mathrm{conf}}(r,s) \le \Delta,
\end{equation}
The consequence of a failed conjunct is scoped to the requested capability, never to the entity's standing.

\gnote{SUGGEST EQUATION: type the act itself, $a = (\text{capability}, \text{constellation}, \alpha, t)$, so that \cref{eq:demand}, \cref{eq:pass}, and the admissibility grammar share one signature; the formalised prose is what downstream agents will consume.}

The freshness clause of \cref{eq:pass} separates two states a single verification flag fuses: \emph{valid} and \emph{eligible}. An artifact is valid when it is genuine, unexpired on its issuer's clock, and its confirmation event real; it is eligible at a gate when, in addition, the confirmation falls inside that gate's freshness window $\Delta$. A government document notarised eight months ago is valid in every sense and still ineligible at an account-opening gate whose law demands a verification no older than six months; a date of birth confirmed twenty years ago stays eligible everywhere, because the fact does not expire and no window is declared against it. Eligibility is a relying-party predicate over coordinate and freshness together, never a flag stored on the cell: the cell records when each confirmation happened, and each gate decides what counts as fresh enough.

This yields a further distinction between evidence disclosure and assurance disclosure. A capability gate receives the signed package needed to satisfy its declared requirement and the lower-tier inclusions on which that package depends, using holder-presented credential and proof mechanisms rather than a bulk disclosure record.\cite{w3c2025vcdm,w3c2025vcdataint,openid2025openid4vp} If a gate requires \texttt{EAID-N4-S5}, the relying service receives the N4-S5 presentation, its provenance, freshness, validity, source class, and the included lower assertions needed to make N4 meaningful. It does not receive higher-tier evidence merely because the actor holds it, because the gate has declared no legal or functional reason for that higher disclosure.

An actor may nevertheless choose to disclose an assurance ceiling. The ceiling statement says that a stronger reusable assurance state exists, for example \texttt{EAID-N10-S7}, while the signed evidence package remains scoped to the lower gate. The ceiling is a voluntary assurance attestation: the actor can signal that a deeper identity state is available, while the relying party obtains only the information justified by the capability it requested. Selective-disclosure cryptographic techniques can support this split, subject to the privacy and correlation limits of the chosen proof format.\cite{w3c2026vcdibbs,openid2025openid4vp} This preserves the practical value of a portable high-assurance signal without collapsing selective disclosure back into full identity exposure.

Anti-money-laundering due diligence is treated as a regulatory mapping over this grammar. Current EU AML law structures third-party reliance through AMLD4, while Regulation (EU) 2024/1624 supplies the incoming directly applicable AML rulebook, including customer due diligence measures, simplified due diligence in lower-risk cases, and enhanced due diligence in higher-risk cases.\cite[arts.~25--27]{eu2015amld4}\cite[arts.~19, 20, 33, and 34]{eu2024amlr} Where AML law applies, obliged entities must satisfy those measures under the law applicable at the relevant time. Their conceptual starting point is the AML relationship between an obliged entity and a customer, including beneficial-owner and risk-factor analysis. The proposed assurance layer starts one level higher: with the legally relevant capability, the actor constellation that seeks to exercise it, and the evidence needed to make the resulting operation imputable.

Regulatory mapping translates a sectoral legal demand into gate predicates while preserving the general ontology. An AML asset transfer may map to simplified, standard, or enhanced due diligence requirements. A high-value contract may require proof of age, legal capacity, authority, and jurisdiction before a natural actor or a J+N constellation can sign. A minor may be identifiable as N while lacking capacity for the requested transaction, so the gate must protect the minor by withholding that capability or routing it through an allowed legal representative constellation. A fraud allegation between two natural actors may require evidence for police, court, or civil recovery even if the original communication gate was low risk. A machine-generated asset sale may require the system to preserve who or what acted, when, under which mandate, with which result, and which N, J+N, or J+N+M constellation supplies the point of imputation.

The result is a two-layer design. Legal identity assurance stores reusable actor and evidence state, while regulatory mappings consume that state through declared gate predicates. AML SDD/CDD/EDD is one such mapping. Contract capacity, minor protection, mandate verification, organizational authority, machine delegation, consumer protection, and evidentiary preservation are additional mappings over the same assurance state.

A further design question is how the assurance state should represent source independence. The model should avoid a general trust score, because low assurance can be legitimate where a low-consequence gate is requested. The more precise concept is attribution resilience: the degree to which independent, source-diverse evidence can sustain attribution, recourse, and regulatory satisfaction if a consequential act is later disputed. Multiple passports, government identity documents, regulated bank relationships, verified addresses, and jurisdictional anchors do not make an actor morally more trustworthy. They make the submitted legal identity state harder to fabricate in full, easier to cross-check for inconsistency, and more useful when lawful resolution later requires an accountable point.

This matters because present onboarding and platform processes often collect either too much weak information or too little legally useful information. A business-bank onboarding process may request public social-media traces that are cheap to create and weakly connected to legal identity, while stronger registry, representative, beneficial-owner, address, and bank-control evidence remains slow, duplicated, and opaque. A platform-mediated rental dispute may process a legally relevant conflict while the claimant has no usable route to identify whether the counterparty is a private host, a commercial operator, or a juridical actor behind the listing. In both cases the problem is a missing regulatory mapping from the requested capability or dispute path to the minimum evidence package needed for compliance, proportionality, and recourse.

The proposed architecture can reserve this without fixing the later scoring mechanism. Gates consume declared predicates; coordinates express disclosure scope and confirmation source; jurisdictional instances hold the relevant legal anchors; and signed legal-identity presentations disclose only the set needed for the requested gate. A later implementation may derive regulatory mapping sets from the same state, such as a bank-onboarding set, a contract-signing set, a minor-protection set, or a platform-dispute set. The principle for the present paper is narrower: source diversity and triangulation increase assurance and attribution resilience, while capability gates prevent low-assurance actors from exercising capabilities for which the evidentiary basis is insufficient.

\subsection{Comparative design evaluation}
The evaluation is analytical, as appropriate for a design-science construct at specification stage. Each comparator is tested against the six requirements fixed in \cref{tab:requirements}. The comparator set follows the alternatives introduced in the problem and related-work sections: flat maximum verification as the control condition, per-credential level-of-assurance frameworks, institutional reusable-KYC reliance, and the proposed grid with confirmation and reliance events.\cite{eu2024amlr,fatf2025recommendations,eu2014eidas,eu2015eidaslevels,nist80063r4,nist80063a4,swift2026kycregistry,biia2020invidem,w3c2025vcdm} The matrix reports design capacity, not implementation performance: it asks what each architecture can express before local policy parameters are chosen. The scoring rule is primitive based: high means the requirement is represented by native state or event fields; partial means part of the requirement is represented while another part remains external policy or process; low means the architecture lacks a native expression for the requirement. \Cref{tab:comparative-evaluation} applies that rule across the four comparators, making the later scenario readings traceable to the six requirements rather than to an impressionistic verdict.

\begin{table}[!htbp]
\caption{Comparative evaluation against the six requirements. Scores report architectural support: low means the design expresses the requirement only through external policy or manual process; partial means it expresses part of the requirement; high means the requirement is native to the model.}
\label{tab:comparative-evaluation}
\centering
\scriptsize
\renewcommand{\arraystretch}{1.18}
\begin{tabularx}{\textwidth}{@{}>{\raggedright\arraybackslash}p{0.17\textwidth}>{\raggedright\arraybackslash}X>{\raggedright\arraybackslash}X>{\raggedright\arraybackslash}X>{\raggedright\arraybackslash}X@{}}
\toprule
Requirement & Flat maximum verification & Per-credential LoA & Institutional reusable KYC & Proposed grid\\
\midrule
R1 Proportional demand & Low: maximal entry bar. & Partial: relying party chooses a level. & Partial: sectoral risk bands. & High: each gate declares the cells it consumes.\\
R2 Assurance/capability separation & Low: verification bit and admission merge. & Partial: level remains credential or scheme bound. & Partial: reuse serves the institution's onboarding purpose. & High: state, demand set, and pass predicate remain distinct.\\
R3 N/J/M imputation & Low: natural-customer template dominates. & Partial: natural-person credentials lead; delegation sits outside the level. & Partial: juridical and beneficial-owner evidence fit AML reuse. & High: N, J, M, and composites share one typed grammar.\\
R4 Granular disclosure & Low: evidence is collected as a package. & Partial: credential presentation can disclose attributes, while level still bundles assurance. & Partial: document exchange remains package oriented. & High: row-addressed assertions support scoped presentations.\\
R5 Jurisdictional continuity & Low: account anchor fixed at enrolment. & Partial: scheme portability exists inside trust frameworks. & Partial: reuse follows participating institutions and current files. & High: jurisdictional anchor and bitemporal event state are stored.\\
R6 Allocated reliance & Low: reliance is an enrolment fact. & Partial: assertion consumption is federated, liability allocation is external. & Partial: third-party reliance exists with retained responsibility. & High: reliance event records gate, parties, law, time, and liability boundary.\\
\bottomrule
\end{tabularx}
\end{table}

The same result can be checked by six small scenario readings. For R1, a low-consequence forum reply needs status and a reachable handle, while an asset transfer needs stronger source classes and a richer constellation; only the grid makes those two demands addresses over the same state. For R2, the same actor may hold a high-assurance passport cell and still fail a child-safety, mandate, payment, or freshness gate; the grid reports that as a gate outcome rather than as a change in identity standing. For R3, a machine-generated offer is admissible only through the attachment path in M0 and the applicable N, J, or composite relation, which lets the architecture model technical action while preserving legal imputation. For R4, an age gate can request N1 at the required source class without receiving address, document scan, or biometric rows. For R5, a move across jurisdictions changes the active anchor for later acts while preserving the past anchor for replay. For R6, a relying service can consume a prior confirmation, yet the reliance event still records who relied, on which package, under which rule version, and within which liability boundary.

\Cref{tab:comparative-evaluation} gives two results. First, flat maximum verification and per-credential LoA designs are coherent baselines, but each is a fold of the grid named in \cref{tab:fold-kill}. The loss appears exactly where a requirement needs cell addressability, event timing, or a distinction between assurance state and capability permission. Second, institutional reusable-KYC reliance already supports part of R6 and some juridical due-diligence work, which is why it scores better than a simple verification bit on reuse. Its object remains a sectoral due-diligence file rather than a portable legal identity-assurance state with typed actors, jurisdictional anchors, and gate-scoped reliance. The proposed artifact is the only evaluated design whose primitives correspond one-to-one to all six requirements.

\section{Discussion}

\subsection{Design rationale and trade-off structure}
The evaluation shows why the artifact is organised as a grid rather than as a ladder. The usual pressure in identity systems is to simplify state until a verifier can read one signal. That pressure is administratively attractive, but \cref{tab:fold-kill} and \cref{tab:comparative-evaluation} show its price: every one-signal design has to move some distinction out of the architecture and into policy prose, manual exception handling, or institutional trust. The proposed grid keeps those distinctions in the address space. This is the central design rationale, even more than the number of source classes or assertion rows. The trade-off map in \cref{tab:discussion-tensions} compresses that discussion into five recurring tensions, showing how the artifact moves each tension into addressable state, declared gates, or reliance records.

\begin{table}[!htbp]
\caption{Design tensions and artifact responses. The responses interpret the evaluated requirements rather than adding new requirements.}
\label{tab:discussion-tensions}
\centering
\footnotesize
\renewcommand{\arraystretch}{1.15}
\begin{tabularx}{\textwidth}{@{}>{\raggedright\arraybackslash}p{0.27\textwidth}>{\raggedright\arraybackslash}p{0.34\textwidth}>{\raggedright\arraybackslash}X@{}}
\toprule
Tension & Flattened response & Artifact response\\
\midrule
Minimisation and recourse & collect a broad package early & address each assertion row and preserve reliance events\\
Inclusion and high-consequence control & impose a high entry bar & admit the status floor and gate later capabilities\\
Reuse and retained responsibility & copy due-diligence files or trust an institutional assertion & record the package, parties, rule version, and liability boundary at reliance\\
Portability and local law & bind the account to one enrolment jurisdiction & store facts once and evaluate jurisdiction-relative predicates at act time\\
Automation and legal imputation & treat machine output as a system event beside the actor model & route machine capability through M0 attachment and admissible constellations\\
\bottomrule
\end{tabularx}
\end{table}

The liveness of the model lies in this relocation of change. The assurance state can accumulate confirmations over time, while each gate remains free to change its predicate when law, risk, institutional policy, or the requested capability changes. A service does not need to decide once that an actor is verified. It declares the cells, source classes, freshness windows, constellation, and jurisdictional anchor that a concrete act requires. The same actor can pass a forum gate, fail a payment gate, pass a contract gate through J+N, and later replay the relied-on state for a dispute without any contradiction in identity standing.

This also explains the cumulative-disclosure claim. The artifact is cumulative for the holder and selective for each relying party. Confirmations gathered for one act can remain available for later gates, but later gates receive only the cells they can justify. In a per-service checklist model, the same documents are repeatedly requested, copied, interpreted, and stored by each institution. In the proposed model, confirmation is reusable and reliance is newly recorded. That distinction keeps reuse from becoming responsibility laundering: the relying party can consume prior confirmation, yet its own reliance event states the decision it made and the legal boundary within which it made it.

\subsection{Institutional implications and adoption conditions}
For relying services, the immediate implication is that identity demand becomes publishable. A service can state a gate profile for a capability: the entity types admitted, the required assertion cells, the minimum source classes, the freshness windows, the jurisdictional anchor, and the reliance event it will create. This turns identity collection from a hidden checklist into a comparable demand. Two services that request the same capability can expose whether one demands more data, fresher evidence, stronger sources, or a broader liability boundary. That comparability is the institutional form of proportionality.

For public authorities, qualified trust-service providers, notaries, banks, registers, auditors, and other high-accountability confirmers, the implication is a shift from document production to confirmation-event production. The institutional act remains familiar: issue an identity document, confirm a register fact, witness a deed, verify a bank relationship, audit a machine artifact, or attest a role. The assurance layer changes the output format. It asks the confirmer to bind the act to explicit rows, source class, valid time, record time, policy version, status channel, and proof material. Verifiable-credential standards can carry much of that substrate, while legal identity assurance supplies the row, source, jurisdiction, and reliance semantics around it.\cite{w3c2025vcdm,w3c2025vcdataint,w3c2025vcbitstringstatus,openid2025openid4vp}

For regulators, the artifact makes over-identification and under-identification visible in the same grammar. A child-safety gate, an AML onboarding gate, a corporate-signing gate, a consumer-refund gate, and a machine-delegation gate can all be expressed as predicates over one state. This does not make the predicates identical. It makes their demands auditable and comparable. The regulator can ask whether a gate requests cells that the act justifies, whether a lower-disclosure path exists where the law permits one, whether a high-consequence act has a lawful point of imputation, and whether a later dispute can replay the relied-on state under the rule version in force at the time.

The adoption condition is network formation around gate profiles and confirmation events. The artifact becomes useful when enough relying services publish gate predicates, enough confirmers issue event-bound attestations, and enough wallets or repositories can present scoped packages. This is a standards and governance problem as much as an engineering problem. The architecture reduces repeated collection only if institutions agree on row identifiers, source-class mappings, status channels, proof formats, and role-equivalence tables across jurisdictions. Machine actors add one more adoption condition: operators must expose M0 attachment and mandate evidence before automated capability can be accepted as legally meaningful.

\section{Conclusion}
This paper proposed progressive legal identity assurance as a reusable assurance layer for digital acts. Its starting point was the mismatch between flat maximum verification, per-credential assurance levels, institutional reusable KYC, and the actual variety of legally relevant acts. The contribution is an artifact that separates entity, actor, and legal identity; represents assurance as a two-axis grid of assertion rows and source classes; treats jurisdiction as an act-time and time-indexed attribute; and records confirmation and reliance as events with provenance, policy version, parties, timing, and liability boundary.

The comparative evaluation supports the design claim. Flat maximum verification, per-credential LoA, and institutional reusable KYC each express important parts of the problem space, but each folds away at least one distinction needed by the six requirements. The proposed grid is the evaluated design whose primitives map directly to proportional demand, assurance/capability separation, N/J/M imputation, granular disclosure, jurisdictional continuity, and allocated reliance. The tier count remains a design parameter. The architectural claim is the separation of assurance state from capability gates, and the preservation of the cell and event structure that makes that separation usable.

\section{Limitations and Future Research}
The first limitation is governance. The model relies on stable assertion-row identifiers, source-class mappings, status channels, proof formats, and role-equivalence tables across jurisdictions. These are standardisation tasks, and their quality will determine whether the architecture remains interoperable or fragments into local dialects. In particular, S6/S7 public, qualified, notarial, and functionally equivalent authority roles need jurisdiction-specific mapping before a production system can treat them as comparable.

The second limitation is strategic behaviour. A coordinate system can be gamed if actors learn which cells unlock which capabilities and can cheaply manufacture weak confirmations. The model addresses this structurally through source classes, binding, confirmation-event correlation, freshness windows, and the confirmation index, but parameter setting remains a governance and risk-analysis task. A reference implementation should test whether gate profiles invite over-requesting by relying services or source shopping by actors.

The third limitation is cost. The model deliberately avoids a cheap confirmed-age shortcut: if a gate needs legally reliable age, it must obtain a confirmation strong enough for that act, even where a weaker self-declaration would be easier. The same cost appears for high-value jurisdictional predicates, beneficial-ownership paths, machine mandates, and notarial or public confirmations. This is a design choice in favour of auditability and recourse, but deployment must still decide where cost, friction, and risk make a lower gate appropriate.

The fourth limitation is scope. The paper specifies the assurance coordinate, actor grammar, gate semantics, and reliance event, but it does not deliver a reference implementation, operational governance process, or empirical pilot. It also leaves aggregation, consensus, and mandate mechanics to the companion papers. Future work should implement the coordinate and event schema, test gate profiles in representative use cases, evaluate privacy and correlation risks of selective-disclosure formats, and measure how well institutions can issue and consume confirmation events without rebuilding per-service document collection.

\bigskip


\bigskip

\printbibliography

@misc{eu2024amlr,
  author    = {{European Parliament and Council of the European Union}},
  title     = {Regulation (EU) 2024/1624 of the European Parliament and of the Council of 31 May 2024 on the prevention of the use of the financial system for the purposes of money laundering or terrorist financing},
  year      = {2024},
  date      = {2024-05-31},
  url       = {https://eur-lex.europa.eu/eli/reg/2024/1624},
  urldate   = {2026-06-08},
}

@misc{eu2015amld4,
  author    = {{European Parliament and Council of the European Union}},
  title     = {Directive (EU) 2015/849 of the European Parliament and of the Council of 20 May 2015 on the prevention of the use of the financial system for the purposes of money laundering or terrorist financing},
  year      = {2015},
  date      = {2015-05-20},
  url       = {https://eur-lex.europa.eu/eli/dir/2015/849/oj/eng},
  urldate   = {2026-07-24},
}

@misc{eu2016gdpr,
  author    = {{European Parliament and Council of the European Union}},
  title     = {Regulation (EU) 2016/679 of the European Parliament and of the Council of 27 April 2016 on the protection of natural persons with regard to the processing of personal data and on the free movement of such data, and repealing Directive 95/46/EC},
  year      = {2016},
  date      = {2016-04-27},
  url       = {https://eur-lex.europa.eu/eli/reg/2016/679/oj/eng},
  urldate   = {2026-06-08},
}

@misc{eu2024aiact,
  author    = {{European Parliament and Council of the European Union}},
  title     = {Regulation (EU) 2024/1689 of the European Parliament and of the Council of 13 June 2024 laying down harmonised rules on artificial intelligence},
  year      = {2024},
  date      = {2024-06-13},
  url       = {https://eur-lex.europa.eu/eli/reg/2024/1689/oj/eng},
  urldate   = {2026-07-24},
}

@misc{eu2024pld,
  author    = {{European Parliament and Council of the European Union}},
  title     = {Directive (EU) 2024/2853 of the European Parliament and of the Council of 23 October 2024 on liability for defective products and repealing Council Directive 85/374/EEC},
  year      = {2024},
  date      = {2024-10-23},
  url       = {https://eur-lex.europa.eu/eli/dir/2024/2853/oj/eng},
  urldate   = {2026-07-24},
}

@misc{ep2020ailiability,
  author    = {{European Parliament}},
  title     = {European Parliament resolution of 20 October 2020 with recommendations to the Commission on a civil liability regime for artificial intelligence (2020/2014(INL))},
  year      = {2020},
  date      = {2020-10-20},
  url       = {https://eur-lex.europa.eu/legal-content/EN/TXT/?uri=oj:JOC_2021_404_R_0006},
  urldate   = {2026-07-24},
}

@misc{cjeu2022luxembourgbusinessregisters,
  author    = {{Court of Justice of the European Union}},
  title     = {Judgment of the Court (Grand Chamber) of 22 November 2022, WM and Sovim SA v Luxembourg Business Registers, Joined Cases C-37/20 and C-601/20},
  year      = {2022},
  date      = {2022-11-22},
  url       = {https://eur-lex.europa.eu/legal-content/EN/TXT/?uri=celex:62020CJ0037},
  urldate   = {2026-07-24},
}

@misc{eu2008romei,
  author    = {{European Parliament and Council of the European Union}},
  title     = {Regulation (EC) No 593/2008 of the European Parliament and of the Council of 17 June 2008 on the law applicable to contractual obligations (Rome I)},
  year      = {2008},
  date      = {2008-06-17},
  url       = {https://eur-lex.europa.eu/eli/reg/2008/593/oj/eng},
  urldate   = {2026-07-24},
}

@misc{eu2007romeii,
  author    = {{European Parliament and Council of the European Union}},
  title     = {Regulation (EC) No 864/2007 of the European Parliament and of the Council of 11 July 2007 on the law applicable to non-contractual obligations (Rome II)},
  year      = {2007},
  date      = {2007-07-11},
  url       = {https://eur-lex.europa.eu/eli/reg/2007/864/oj/eng},
  urldate   = {2026-07-24},
}

@misc{eu2000ecommerce,
  author    = {{European Parliament and Council of the European Union}},
  title     = {Directive 2000/31/EC of the European Parliament and of the Council of 8 June 2000 on certain legal aspects of information society services, in particular electronic commerce, in the Internal Market},
  year      = {2000},
  date      = {2000-06-08},
  url       = {https://eur-lex.europa.eu/eli/dir/2000/31/oj/eng},
  urldate   = {2026-07-24},
}

@misc{eu2024amld6,
  author    = {{European Parliament and Council of the European Union}},
  title     = {Directive (EU) 2024/1640 of the European Parliament and of the Council of 31 May 2024 on the mechanisms to be put in place by Member States for the prevention of the use of the financial system for the purposes of money laundering or terrorist financing},
  year      = {2024},
  date      = {2024-05-31},
  url       = {https://eur-lex.europa.eu/eli/dir/2024/1640/oj/eng},
  urldate   = {2026-07-24},
}

@misc{eu2017companylaw,
  author    = {{European Parliament and Council of the European Union}},
  title     = {Directive (EU) 2017/1132 of the European Parliament and of the Council of 14 June 2017 relating to certain aspects of company law},
  year      = {2017},
  date      = {2017-06-14},
  url       = {https://eur-lex.europa.eu/eli/dir/2017/1132/oj/eng},
  urldate   = {2026-07-24},
}

@misc{uncitral2005electroniccommunications,
  author    = {{United Nations Commission on International Trade Law}},
  title     = {United Nations Convention on the Use of Electronic Communications in International Contracts},
  year      = {2005},
  date      = {2005-11-23},
  url       = {https://uncitral.un.org/en/texts/ecommerce/conventions/electronic_communications},
  urldate   = {2026-07-24},
}

@misc{uncitral2022mlit,
  author    = {{United Nations Commission on International Trade Law}},
  title     = {Model Law on the Use and Cross-border Recognition of Identity Management and Trust Services},
  year      = {2022},
  date      = {2022},
  url       = {https://uncitral.un.org/en/mlit},
  urldate   = {2026-07-24},
}

@misc{uncitral2024automatedcontracting,
  author    = {{United Nations Commission on International Trade Law}},
  title     = {Model Law on Automated Contracting},
  year      = {2024},
  date      = {2024},
  url       = {https://uncitral.un.org/en/mlac},
  urldate   = {2026-07-24},
}

@misc{fatf2023beneficialownershiplegalpersons,
  author    = {{Financial Action Task Force}},
  title     = {Guidance on Beneficial Ownership of Legal Persons},
  year      = {2023},
  date      = {2023-03},
  url       = {https://www.fatf-gafi.org/en/publications/Fatfrecommendations/Guidance-Beneficial-Ownership-Legal-Persons.html},
  urldate   = {2026-07-24},
}

@misc{fatf2024beneficialownershiplegalarrangements,
  author    = {{Financial Action Task Force}},
  title     = {Guidance on Beneficial Ownership and Transparency of Legal Arrangements},
  year      = {2024},
  date      = {2024-03},
  url       = {https://www.fatf-gafi.org/en/publications/Fatfrecommendations/Guidance-Beneficial-Ownership-Transparency-Legal-Arrangements.html},
  urldate   = {2026-07-24},
}

@misc{fatf2020digitalidentity,
  author    = {{Financial Action Task Force}},
  title     = {Guidance on Digital Identity},
  year      = {2020},
  date      = {2020-03},
  url       = {https://www.fatf-gafi.org/en/publications/Financialinclusionandnpoissues/Digital-identity-guidance.html},
  urldate   = {2026-07-24},
}

@techreport{enisa2024ridpgoodpractices,
  author      = {{European Union Agency for Cybersecurity}},
  title       = {Remote ID Proofing: Good Practices},
  institution = {ENISA},
  year        = {2024},
  date        = {2024-03-12},
  url         = {https://www.enisa.europa.eu/publications/remote-id-proofing-good-practices},
  urldate     = {2026-07-24},
}

@techreport{etsi2025ts119461,
  author      = {{European Telecommunications Standards Institute}},
  title       = {{ETSI} TS 119 461 V2.1.1: Electronic Signatures and Trust Infrastructures; Policy and Security Requirements for Trust Service Components Providing Identity Proofing of Trust Service Subjects},
  institution = {ETSI},
  type        = {Technical Specification},
  year        = {2025},
  date        = {2025-02},
  version     = {2.1.1},
  url         = {https://www.etsi.org/deliver/etsi_ts/119400_119499/119461/02.01.01_60/ts_119461v020101p.pdf},
  urldate     = {2026-07-24},
}

@misc{openid2025openid4vp,
  author    = {{OpenID Foundation}},
  title     = {{OpenID} for Verifiable Presentations 1.0},
  year      = {2025},
  date      = {2025-07-09},
  url       = {https://openid.net/specs/openid-4-verifiable-presentations-1_0-final.html},
  urldate   = {2026-07-24},
}

@techreport{w3c2025vcdataint,
  author      = {{World Wide Web Consortium}},
  title       = {Verifiable Credential Data Integrity 1.0: Securing the Integrity of Verifiable Credential Data},
  institution = {W3C},
  type        = {Recommendation},
  year        = {2025},
  date        = {2025-05-15},
  url         = {https://www.w3.org/TR/vc-data-integrity/},
  urldate     = {2026-07-24},
}

@techreport{w3c2025vcbitstringstatus,
  author      = {{World Wide Web Consortium}},
  title       = {Bitstring Status List v1.0: Privacy-preserving Status Information for Verifiable Credentials},
  institution = {W3C},
  type        = {Recommendation},
  year        = {2025},
  date        = {2025-05-15},
  url         = {https://www.w3.org/TR/vc-bitstring-status-list/},
  urldate     = {2026-07-24},
}

@techreport{w3c2026vcdibbs,
  author      = {{World Wide Web Consortium}},
  title       = {Data Integrity BBS Cryptosuites v1.0},
  institution = {W3C},
  type        = {Candidate Recommendation Draft},
  year        = {2026},
  date        = {2026-04-07},
  url         = {https://www.w3.org/TR/vc-di-bbs/},
  urldate     = {2026-07-24},
}

@misc{edpb2020dpbdd,
  author    = {{European Data Protection Board}},
  title     = {Guidelines 4/2019 on Article 25 Data Protection by Design and by Default},
  year      = {2020},
  date      = {2020-10-20},
  url       = {https://www.edpb.europa.eu/documents/guideline/guidelines-42019-on-article-25-data-protection-by-design-and-by-default_en},
  urldate   = {2026-07-24},
}

@article{snodgrass1992temporaldatabaseglossary,
  author   = {Jensen, Christian S. and Clifford, James and Gadia, Shashi K. and Segev, Arie and Snodgrass, Richard T.},
  title    = {A Glossary of Temporal Database Concepts},
  journal  = {SIGMOD Record},
  year     = {1992},
  volume   = {21},
  number   = {3},
  pages    = {35--43},
  doi      = {10.1145/140979.140996},
  url      = {https://sigmodrecord.org/publications/sigmodRecord/9209/pdfs/140979.140996.pdf},
  urldate  = {2026-07-24},
}

@misc{hcch1961apostille,
  author    = {{Hague Conference on Private International Law}},
  title     = {Convention of 5 October 1961 Abolishing the Requirement of Legalisation for Foreign Public Documents},
  year      = {1961},
  date      = {1961-10-05},
  url       = {https://www.hcch.net/en/instruments/conventions/full-text/?cid=41},
  urldate   = {2026-07-24},
}

@misc{tesco1971nattrass,
  author    = {{United Kingdom House of Lords}},
  title     = {Tesco Supermarkets Ltd v Nattrass [1971] UKHL 1},
  year      = {1971},
  date      = {1971-03-31},
  url       = {https://www.bailii.org/uk/cases/UKHL/1971/1.html},
  urldate   = {2026-07-24},
}

@misc{wyoming2021daosupplement,
  author    = {{Wyoming Legislature}},
  title     = {Senate File No. SF0038: Decentralized Autonomous Organizations Supplement},
  year      = {2021},
  date      = {2021},
  url       = {https://www.wyoleg.gov/Legislation/2021/SF0038},
  urldate   = {2026-07-24},
}

@misc{wyoming2024duna,
  author    = {{Wyoming Legislature}},
  title     = {Senate File No. SF0050: Decentralized Unincorporated Nonprofit Association Act},
  year      = {2024},
  date      = {2024},
  url       = {https://www.wyoleg.gov/Legislation/2024/SF0050},
  urldate   = {2026-07-24},
}

@misc{cftc2023ookidao,
  author    = {{Commodity Futures Trading Commission}},
  title     = {Statement of CFTC Division of Enforcement Director Ian McGinley on the Ooki DAO Litigation Victory},
  year      = {2023},
  date      = {2023-06-09},
  url       = {https://www.cftc.gov/PressRoom/PressReleases/8715-23},
  urldate   = {2026-07-24},
}

@book{maturana1980autopoiesis,
  author    = {Maturana, Humberto R. and Varela, Francisco J.},
  title     = {Autopoiesis and Cognition: The Realization of the Living},
  publisher = {D. Reidel Publishing Company},
  location  = {Dordrecht},
  year      = {1980},
  series    = {Boston Studies in the Philosophy and History of Science},
  volume    = {42},
  doi       = {10.1007/978-94-009-8947-4},
  url       = {https://link.springer.com/book/10.1007/978-94-009-8947-4},
  urldate   = {2026-06-08},
}

@book{luhmann1995social,
  author     = {Luhmann, Niklas},
  title      = {Social Systems},
  publisher  = {Stanford University Press},
  location   = {Stanford, CA},
  year       = {1995},
  translator = {Bednarz, John and Baecker, Dirk},
  isbn       = {0804726256},
}

@book{luhmann2004law,
  author     = {Luhmann, Niklas},
  title      = {Law as a Social System},
  publisher  = {Oxford University Press},
  location   = {Oxford and New York},
  year       = {2004},
  translator = {Ziegert, Klaus A.},
  editor     = {Kastner, Fatima and Nobles, Richard and Schiff, David and Ziegert, Rosamund},
  isbn       = {0198262388},
  url        = {https://academic.oup.com/book/50417},
  urldate    = {2026-06-08},
}

@misc{eu2014eidas,
  author    = {{European Parliament and Council of the European Union}},
  title     = {Regulation (EU) No 910/2014 of the European Parliament and of the Council of 23 July 2014 on electronic identification and trust services for electronic transactions in the internal market},
  year      = {2014},
  date      = {2014-07-23},
  url       = {https://eur-lex.europa.eu/eli/reg/2014/910/oj},
  urldate   = {2026-06-08},
}

@misc{eu2015eidaslevels,
  author    = {{European Commission}},
  title     = {Commission Implementing Regulation (EU) 2015/1502 of 8 September 2015 on setting out minimum technical specifications and procedures for assurance levels for electronic identification means},
  year      = {2015},
  date      = {2015-09-08},
  url       = {https://eur-lex.europa.eu/eli/reg_impl/2015/1502/oj},
  urldate   = {2026-06-08},
}

@misc{eu2024eidas2,
  author    = {{European Parliament and Council of the European Union}},
  title     = {Regulation (EU) 2024/1183 of the European Parliament and of the Council of 11 April 2024 amending Regulation (EU) No 910/2014 as regards establishing the European Digital Identity Framework},
  year      = {2024},
  date      = {2024-04-11},
  url       = {https://eur-lex.europa.eu/eli/reg/2024/1183/oj},
  urldate   = {2026-06-08},
}

@misc{nist80063r4,
  author    = {{National Institute of Standards and Technology}},
  title     = {Digital Identity Guidelines},
  year      = {2025},
  date      = {2025-07},
  series    = {NIST Special Publication 800-63-4},
  doi       = {10.6028/NIST.SP.800-63-4},
  url       = {https://pages.nist.gov/800-63-4/sp800-63.html},
  urldate   = {2026-06-08},
}

@misc{nist80063a4,
  author    = {{National Institute of Standards and Technology}},
  title     = {Identity Proofing and Enrollment},
  year      = {2025},
  date      = {2025-07},
  series    = {NIST Special Publication 800-63A-4},
  doi       = {10.6028/NIST.SP.800-63A-4},
  url       = {https://pages.nist.gov/800-63-4/sp800-63a.html},
  urldate   = {2026-06-08},
}

@misc{fatf2025recommendations,
  author    = {{Financial Action Task Force}},
  title     = {International Standards on Combating Money Laundering and the Financing of Terrorism and Proliferation: The FATF Recommendations},
  year      = {2025},
  url       = {https://www.fatf-gafi.org/content/dam/fatf-gafi/recommendations/FATF%20Recommendations%202012.pdf.coredownload.inline.pdf},
  urldate   = {2026-06-08},
}

@article{hevner2004designscience,
  author    = {Hevner, Alan R. and March, Salvatore T. and Park, Jinsoo and Ram, Sudha},
  title     = {Design Science in Information Systems Research},
  journal   = {MIS Quarterly},
  year      = {2004},
  volume    = {28},
  number    = {1},
  pages     = {75--105},
  url       = {https://aisel.aisnet.org/misq/vol28/iss1/6/},
  urldate   = {2026-06-08},
}

@book{kelsen1967pure,
  author     = {Kelsen, Hans},
  title      = {Pure Theory of Law},
  translator = {Knight, Max},
  publisher  = {University of California Press},
  location   = {Berkeley},
  year       = {1967},
  url        = {https://www.ucpress.edu/books/pure-theory-of-law/hardcover},
  urldate    = {2026-06-09},
}

@book{popper1972objective,
  author    = {Popper, Karl R.},
  title     = {Objective Knowledge: An Evolutionary Approach},
  publisher = {Clarendon Press},
  location  = {Oxford},
  year      = {1972},
  url       = {https://openlibrary.org/works/OL1984562W/Objective_Knowledge},
  urldate   = {2026-06-09},
}

@book{goffman1967interaction,
  author    = {Goffman, Erving},
  title     = {Interaction Ritual: Essays on Face-to-Face Behavior},
  publisher = {Anchor Books},
  location  = {Garden City, NY},
  year      = {1967},
}

@book{esposito2022artificial,
  author    = {Esposito, Elena},
  title     = {Artificial Communication: How Algorithms Produce Social Intelligence},
  publisher = {MIT Press},
  location  = {Cambridge, MA},
  year      = {2022},
  series    = {Strong Ideas},
  isbn      = {9780262046664},
  url       = {https://artificialcommunication.mitpress.mit.edu/},
  urldate   = {2026-06-09},
}

@misc{swift2026kycregistry,
  author    = {{Society for Worldwide Interbank Financial Telecommunication}},
  title     = {The {KYC} Registry},
  year      = {2026},
  url       = {https://www.swift.com/our-solutions/compliance-and-shared-services/financial-crime-compliance/kyc-registry},
  urldate   = {2026-06-09},
}

@misc{biia2020invidem,
  author    = {{Business Information Industry Association}},
  title     = {Nordic Banks' Agreement on One {KYC} Standard a Unique Advantage for New Utility},
  year      = {2020},
  date      = {2020-08-25},
  url       = {https://www.biia.com/nordic-banks-agreement-on-one-kyc-standard-a-unique-advantage-for-new-utility/},
  urldate   = {2026-06-09},
}

@techreport{w3c2025vcdm,
  author       = {Sporny, Manu and Thibodeau Jr., Ted and Herman, Ivan and Cohen, Gabe and Jones, Michael B.},
  title        = {Verifiable Credentials Data Model v2.0},
  institution  = {World Wide Web Consortium ({W3C})},
  type         = {{W3C} Recommendation},
  year         = {2025},
  date         = {2025-05-15},
  url          = {https://www.w3.org/TR/vc-data-model-2.0/},
  urldate      = {2026-06-09},
}

@article{schardong2022ssi,
  author    = {Schardong, Frederico and Cust{\'o}dio, Ricardo},
  title     = {Self-Sovereign Identity: A Systematic Review, Mapping and Taxonomy},
  journal   = {Sensors},
  year      = {2022},
  volume    = {22},
  number    = {15},
  pages     = {5641},
  doi       = {10.3390/s22155641},
  url       = {https://www.mdpi.com/1424-8220/22/15/5641},
  urldate   = {2026-06-09},
}

@misc{gleif2025lei,
  author    = {{Global Legal Entity Identifier Foundation}},
  title     = {The Legal Entity Identifier ({LEI}) and the {ISO}~17442 Standard},
  year      = {2025},
  url       = {https://www.gleif.org/en/about-lei/iso-17442-the-lei-code-structure},
  urldate   = {2026-06-09},
}

@misc{gleif2024vlei,
  author    = {{Global Legal Entity Identifier Foundation}},
  title     = {The Verifiable {LEI} (v{LEI})},
  year      = {2024},
  url       = {https://www.gleif.org/en/organizational-identity/introducing-the-verifiable-lei-vlei},
  urldate   = {2026-06-09},
}

@techreport{eudiarf2026,
  author      = {{European Commission}},
  title       = {The European Digital Identity Wallet Architecture and Reference Framework},
  institution = {European Commission, European Digital Identity Cooperation Group ({EDICG})},
  type        = {Architecture and Reference Framework},
  year        = {2026},
  version      = {3.0.0},
  url         = {https://github.com/eu-digital-identity-wallet/eudi-doc-architecture-and-reference-framework},
  urldate     = {2026-07-24},
}

@misc{eudiwallet2026ios,
  author       = {{European Commission}},
  title        = {{EUDI} Wallet Reference Implementation: i{OS} Wallet User Interface},
  howpublished = {Open-source software repository, eu-digital-identity-wallet},
  year         = {2026},
  url          = {https://github.com/eu-digital-identity-wallet/eudi-app-ios-wallet-ui},
  urldate      = {2026-06-09},
}

@misc{ch2024eid,
  author    = {{Swiss Confederation}},
  title     = {Federal Act on Electronic Identity and Other Electronic Credentials (E-{ID} Act)},
  year      = {2024},
  date      = {2024-12-20},
  url       = {https://www.eid.admin.ch/en/parlament-verabschiedet-das-e-id-gesetz-e},
  urldate   = {2026-06-09},
}

@misc{eidas2concerns2023,
  author    = {{Digital Watch Observatory}},
  title     = {Concerns Raised over {EU} Digital {ID} Wallet's Impact on Privacy and Discrimination},
  year      = {2023},
  date      = {2023-06-21},
  url       = {https://dig.watch/updates/concerns-raised-over-eu-digital-id-wallets-impact-on-privacy-and-discrimination},
  urldate   = {2026-06-09},
}

\swDeclarationsPage

\end{document}